\documentclass[12pt]{article}

\usepackage{amssymb,amsfonts,amsthm}
\usepackage[pdftex]{graphicx}
\usepackage{subfigure}
\usepackage{amssymb}
\usepackage{amsthm}
\usepackage{amsmath}
\usepackage{bm}             
\usepackage[mathscr]{eucal}

\usepackage{cancel}

\usepackage{psfrag}

\usepackage{multirow}

\usepackage{wasysym}

\usepackage{color}

\def\hsp5{\hspace{5mm}}

\newcommand{\sfrac}[2]{{\textstyle{#1\over#2}}}

\theoremstyle{plain}
\newtheorem{theorem}{Theorem}[section]

\theoremstyle{remark}

\title{\sc Global dynamics and inflationary center manifold and slow-roll approximants}

\begin{document}

\author{{\sc Artur Alho}\thanks{Electronic address:
{\tt aalho@math.ist.utl.pt}} \\[1ex]
Center for Mathematical Analysis, Geometry and Dynamical Systems, \\
Technical University of Lisbon, 1049-001 Lisbon, Portugal
\and \\
{\sc Claes Uggla}\thanks{Electronic address:
{\tt claes.uggla@kau.se}} \\[1ex]
Department of Physics, \\
University of Karlstad, S-651 88 Karlstad, Sweden \\[2ex] }

\date{}
\maketitle

\begin{abstract}

We consider the familiar problem of a minimally coupled scalar field with
quadratic potential in flat Friedmann-Lema\^itre-Robertson-Walker cosmology
to illustrate a number of techniques and tools, which can be applied to a
wide range of scalar field potentials and problems in e.g. modified gravity.
We present a global and regular dynamical systems description that yields a
global understanding of the solution space, including asymptotic features. We
introduce dynamical systems techniques such as center manifold expansions and
use Pad\'e approximants to obtain improved approximations for the
`attractor solution' at early times. We also show that future asymptotic
behavior is associated with a limit cycle, which shows that manifest
self-similarity is asymptotically broken toward the future, and give
approximate expressions for this behavior. We then combine these results to
obtain global approximations for the attractor solution, which, e.g., might
be used in the context of global measures. In addition we elucidate the
connection between slow-roll based approximations and the attractor solution,
and compare these approximations with the center manifold based approximants.

\end{abstract}

\centerline{\bigskip\noindent PACS numbers: 04.20.-q, 98.80.-k, 98.80.Bp,
98.80.Jk}

\section{Introduction}

The present paper is part of a research program that uses global dynamical
systems methods to study spatially homogeneous cosmological models and
cosmological perturbations for a wide variety of sources and gravity theories. The
overall purpose of this program is: (i) To produce regularized global
dynamical systems in order to obtain a global picture of the solution space
of various cosmological models, thus determining the possible and likely
cosmological consequences of different physical assumptions. (ii) To describe
the asymptotic features of the models. (iii) To introduce and assess
different approximation schemes that approximately describe solutions in
their asymptotic regimes, and to combine them to give the entire solutions
approximatively, in global dynamical systems settings.

To achieve the above goals require the introduction of various more or less
sophisticated mathematical techniques in an area that to a large extent is
dominated by heuristic arguments and local considerations. This is the price
to be payed in order to achieve clarified rigor and to obtain a global
understanding of cosmological possibilities that contextualize various
heuristic results when they are essentially correct, and to sometimes reveal
that some heuristic beliefs are actually unfounded myths that rather reflect
hopes than what is actually the case.

In the present paper we consider a scalar field that is minimally coupled to
the Einstein equations in flat Friedmann-Lema\^{i}tre-Robertson-Walker (FLRW)
cosmology with a quadratic potential. This is a simple and thoroughly studied
problem and this is precisely why we choose it; the main purpose with the
present paper is to present new ideas and methods in the setting of this
familiar problem which are subsequently to be used (and sometimes modified)
in a series of forthcoming papers that address increasingly complicated and
therefore less understood problems. Moreover, some of the results for the
current model turn out to reflect some rather general features, valid for
much more general problems. It should, however, be pointed out that
discussions about the recent results of BICEP2~\cite{bicep2} (see
e.g.~\cite{okaetal14,mawan14} as examples of recent reactions to BICEP2) also
motivate taking a renewed look at the classic problem of a minimally coupled
scalar field with a quadratic potential.

The present scalar field problem has a quite long history. For brevity we
only give the key historical references that serve as the starting point for
the present work, namely the papers by Belinski\v{\i} {\it et
al}.~\cite{beletal85a,beletal85b,belkha87,beletal88},
Rendall~\cite{ren02,ren07}, and Liddle {\it et al}.~\cite{lidetal94}. (A few
other examples of references that describe minimally coupled scalar field
cosmology in dynamical systems settings are~\cite{col03,beyesc13,fadetal14},
with additional references therein.)

The first goal of this paper is to present a global dynamical systems picture
of the solution space, which furthermore gives a clear picture of the various
past and future asymptotic states. This naturally brings the so-called
attractor solution into focus in a global dynamical systems setting.

The second goal is to determine possible asymptotic behavior at early and
late times. In particular we establish that late time behavior is associated
with a particular type of asymptotic self-similarity breaking. At early times
we show that the generic asymptotes are given by a massless scalar field
state while a special type of de Sitter state describes the asymptotic
behavior of the so-called attractor solution.

The third goal is to introduce, describe and assess a variety of
approximation techniques that all boil down to giving an approximation for
the attractor solution. In particular, we will show that the familiar
slow-roll approximation and its slow-roll expansion extensions is just such
an approximation scheme for the attractor solution at early times. This
suggests that one should explicitly take the past asymptotic behavior of the
attractor solution as the starting point for an approximation scheme. The
attractor solution turns out to originate from a non-hyperbolic fixed point,
which can be treated by means of center manifold theory to yield an expansion
that gives an approximation for the attractor solution. In addition we show
how so-called Pad{\'e} (or, more generally, Canterbury) approximants can be
used to analytically improve the range and rate of convergence for the
various approximation expansions.

The fourth goal is to combine the center manifold based Pad{\'e} approximants
and the approximate solutions for the non-self-similar behavior at late times
to obtain global approximations for the attractor solution.


The outline of the paper is as follows. In the next section we first present
a regular unconstrained 2-dimensional dynamical system on a compact state
space for a scalar field with a quadratic potential that is minimally coupled
to the Einstein equations in flat FLRW cosmology. We then perform a global
dynamical systems analysis identifying future and past behavior, and it is
shown that the so-called attractor solution corresponds to a 1-dimensional
unstable center submanifold of a certain non-hyperbolic fixed point. In
particular we derive series expansions and improve their convergence
properties and range by using Pad{\'e} approximants. We also give
approximations for the behavior at late times and use this together with the
center manifold results to provide a global analytical approximate
description of the attractor solution. In Section~\ref{sec:slowroll} we
extend the slow-roll approximant results of Liddle {\it et
al}.~\cite{lidetal94} to higher orders, and show how they provide
approximations for the attractor solution in the present global dynamical
systems context. These results, in combination with numerics, are then
compared with the center manifold results to assess the accuracy and range of
the various types of approximations. In Section~\ref{sec:concl} we further
discuss our results and their implications for more general potentials and
models.


%
\section{Dynamical systems approach}\label{sec:dynsysappr}

The Einstein and scalar field equations for a minimally coupled scalar field $\phi(t)$
with potential $V(\phi) = \frac12 m^2 \phi^2$ for flat FLRW cosmology are
given by
\begin{subequations}\label{Hphieq}
\begin{align}
3H^2 &= \sfrac12 \dot{\phi}^2 + \sfrac12 m^2 \phi^2,\label{Gauss1}\\
\dot{H} &= -\sfrac{1}{2} \dot{\phi}^2,\label{Ray1}\\
0 &= \ddot{\phi} + 3H\dot{\phi} + m^2\phi,\label{KG}
\end{align}
\end{subequations}
where an overdot signifies the derivative with respect to synchronous time,
$t$. Throughout we use (reduced Planck) units such that $c=1=8\pi G$, where
$c$ is the speed of light and $G$ is the gravitational constant. (In the
inflationary literature the gravitational constant $G$ is often replaced by
the Planck mass, $G=1/m_\mathrm{Pl}^2$.) In addition, $H$ is the Hubble
variable, which is given by $H = \dot{a}/a$, where $a(t)$ is the cosmological
scale factor, and throughout we assume an expanding Universe, i.e. $H>0$. The
deceleration parameter $q$ is defined via $\dot{H} = -(1 + q)H^2$, and due
to~\eqref{Ray1} it is therefore given by
\begin{equation}\label{qscalar}
q = - 1 + \frac{1}{2}\left(\frac{\dot{\phi}}{H}\right)^2.
\end{equation}
It is also of interest to define the effective equation of state parameter
$w_\phi$:
\begin{equation}\label{wscalar}
w_\phi := \frac{p_\phi}{\rho_\phi} = \frac{\sfrac12 \dot{\phi}^2 - \sfrac12 m^2 \phi^2}{\sfrac12 \dot{\phi}^2 + \sfrac12 m^2 \phi^2},
\end{equation}
and hence $q = \frac12(1+3w_\phi)$, a relation that holds for an arbitrary
potential in flat FLRW cosmology.

Heuristically $3H^2$ in eq.~\eqref{Gauss1} can be viewed as an energy which
according to eq.~\eqref{Ray1} is decreasing. Furthermore, eq.~\eqref{KG} can
be interpreted as describing a nonlinearly damped harmonic oscillator, where
$3H\dot{\phi}$ is the friction term, which suggests that the scalar field
$\phi$ oscillates with decreasing amplitude toward the future and that
$(H,\dot{\phi},V(\phi),\phi)\rightarrow (0,0,0,0)$, which is indeed the case.
However, in these variables $q$ in eq.~\eqref{qscalar} becomes ill-defined,
and the above heuristic picture says nothing about what happens with this
important physical observable.

Moreover, reversing the time direction leads to a heuristic picture of a
harmonic oscillator that gains energy, but in the past limit where
$H\rightarrow \infty$ the energy driving term $3H\dot{\phi}$ becomes
ill-defined and again the heuristic picture breaks down. One might take the
viewpoint that the limit $H\rightarrow \infty$ is irrelevant since it would
take one beyond the Planck regime where the classical description fails, but
this would be a mistake. As we will see, an understanding of the classical
limit $H\rightarrow \infty$ is necessary in order to understand the situation
for large $H$, where the classical picture is expected to hold.

Furthermore, these heuristic considerations only yield rough
\emph{qualitative} pictures, which in addition say little about the
asymptotic regimes. To obtain \emph{quantitative} results, and to obtain
results that yield a global dynamical picture, illustrating the entire
solution space and its properties, requires a reformulation of the field
equations that regularize them on a bounded state space that includes the
limits $H\rightarrow \infty$ and $H\rightarrow 0$.

\subsection{Global dynamical systems formulation}\label{dynsysform}

To obtain a global dynamical system on a relatively compact state space,
which can be regularly extended to include its boundary, we define
\begin{subequations}\label{Tthetadef}
\begin{align}
T &:= \frac{m}{m + H},\\
\theta &:= \arctan\left(\frac{\dot{\phi}}{m \phi}\right),
\end{align}
\end{subequations}
and a new independent variable $\bar{\tau}$,
\begin{equation}\label{bartaudef}
\frac{d\bar{\tau}}{dt} := m + H,
\end{equation}
which leads to the dynamical system
\begin{subequations}\label{dynsys}
\begin{align}
\frac{dT}{d\bar{\tau}} &= 3T(1-T)^2\sin^2\theta = \frac32T(1-T)^2(1 - \cos 2\theta),\label{Teq}\\
\frac{d\theta}{d\bar{\tau}} &= - T - \frac32(1-T)\sin 2\theta .
\end{align}
\end{subequations}

The physical relatively compact state space ${\bf S}$ is defined by a finite
cylinder with $T=0$ and $T=1$ as invariant boundary subsets, which can be
included to yield the state space $\bar{\bf S}$. The interpretation in terms
of a cylinder is perhaps most easily seen by introducing $T$ and the two
variables
\begin{equation}
\Sigma_\dagger = \frac{\dot{\phi}}{\sqrt{6}H} \quad \text{and} \quad X = \frac{m\phi}{\sqrt{6}H}.
\end{equation}
Together with the time choice $\bar{\tau}$, this leads to a 3-dimensional
dynamical system for $T,\Sigma_\dagger,X$ obeying the 
constraint~\eqref{Gauss1}, which takes the form
\begin{equation}\label{Gauss2}
1=\Sigma_\dagger^2 + X^2,
\end{equation}
and thus $T,\Sigma_\dagger$ and $X$ describe a cylinder with unit radius.
Solving the constraint~\eqref{Gauss2} globally by introducing $\theta$ via
\begin{equation}
\Sigma_\dagger = \sin\theta,\quad X = \cos\theta
\end{equation}
yields the present dynamical systems formulation. Note that the
system~\eqref{dynsys} admits a discrete symmetry since it is invariant under
the transformation $\theta \rightarrow \theta + \pi$, which is a reflection
of that $V(\phi)$ is invariant under $\phi \rightarrow - \phi$.

In terms of the new variables the Hubble variable $H$, the scalar field
$\phi$, and $\dot{\phi}$, are given by
\begin{subequations}\label{Newvar}
\begin{align}
\frac{H}{m} &= \frac{1-T}{T},\label{HT}\\
\phi &= \sqrt{6}\left(\frac{1-T}{T}\right)\cos\theta,\\
\dot{\phi} &= \sqrt{6}m\left(\frac{1-T}{T}\right)\sin\theta,
\end{align}
\end{subequations}
and hence constant $T$ surfaces in the state space ${\bf S}$ correspond to
constant values of $H$. The deceleration parameter $q$ and the effective
equation of state parameter $w_\phi$ are given by
\begin{subequations}
\begin{align}
q &= - 1 + 3\sin^2\theta = \sfrac12(1 - 3\cos 2\theta),\\
w_\phi &= 2\sin^2\theta - 1 = - \cos 2\theta.\label{w}
\end{align}
\end{subequations}
As a consequence a solution is accelerating as long as $\sin^2\theta <
\frac13$ (for future use in figures, note that $\arcsin(1/\sqrt{3})\approx
\frac{\pi}{5}$).
%

Note that the present variables are the same as those used by Rendall
in~\cite{ren02}, where $T$ was denoted by $u$ and $\bar{\tau}$ by $\tau$,
although the present state space was never used for global purposes in that
paper. In addition, $T$ and $\theta$ are closely related to the variables
$\rho,\theta,\psi$ used in e.g.~\cite{belkha87}, where $\theta = \pi/4$ in
the flat FLRW case while $\psi$ corresponds to the present $\theta$. However,
that work used different projections to describe the dynamics. In our opinion
the presently used state space description has the advantage of clearly
giving a global description of the dynamics, including at past and future
asymptotic times.

Finally, it should also be pointed out that the system~\eqref{dynsys} is a
\emph{reduced} system. The equation for the scale factor $a$ has been
decoupled. This equation is obtained from $H=\dot{a}/a$, where the above
changes of independent and dependent variables yield
\begin{equation}
\frac{da}{d\bar{\tau}} = (1-T)a,
\end{equation}
which leads to a quadrature for $a$ once the system~\eqref{dynsys} has been
solved. If one furthermore wants to express things in terms of the
synchronous proper time variable $t$, one needs to integrate the relation
$mdt = T(\bar{\tau})d\bar{\tau}$, which follows from eqs.~\eqref{bartaudef}
and~\eqref{HT}.

\subsection{Dynamical system for early times}\label{sec:appendixA}

Although the system~\eqref{dynsys} gives an illustrative global picture of
the solution space, it might not yield the best variables for describing
non-global state space structures. Below we are going to be interested in
approximating the attractor solutions at early times. In this case we obtain
a simpler system, which results in a more convenient analysis, by introducing
the variables
\begin{subequations}\label{dynsystildeT}
\begin{align}
\tilde{T} &:= \frac{m}{H},\\
\theta &:= \arctan\left(\frac{\dot{\phi}}{m \phi}\right),
\end{align}
\end{subequations}
and a new independent variable ${\tau}$,
\begin{equation}
\frac{d{\tau}}{dt} := H,
\end{equation}
which in an inflationary context can be viewed as the number of $e$-folds
$N$, i.e., $\tau=N$. This leads to the (auxiliary) dynamical system
\begin{subequations}\label{dynsys2}
\begin{align}
\frac{d\tilde{T}}{d\tau} &= 3\tilde{T}\sin^2\theta = \frac32\tilde{T}(1 - \cos 2\theta),\label{tTeq}\\
\frac{d\theta}{d\tau} &= - \tilde{T} - \frac32\sin 2\theta ,
\end{align}
\end{subequations}
which can be obtained from~\eqref{dynsys} by taking the small $T$ limit.

That there exist dynamical systems that describe local dynamics more
conveniently than systems that describe the dynamics globally should not come
as a surprise: different regimes induce extra structures which can be used in
dynamical systems formulations. Near initial singularities it is natural to
adapt both dependent and independent variables to the Hubble (equivalently,
the expansion) variable, due to that the Hubble variable provides a natural
scale in this regime, as discussed in e.g.~\cite{ugg13b}. On the other hand,
the oscillatory regime at late times in the present case depends on the
minimum of the potential which is characterized by $\frac{d^2V}{d\phi^2} =
m^2$, and thus $m$ provides the natural scale in the late time regime. These
features are reflected in our choice of global variable $T = m/(H+m)$ and
independent variable $\bar{\tau}$, defined via $d\bar{\tau} = (m + H)dt$,
which incorporate (and interpolate between) the two natural asymptotic
scales.

\subsection{Global dynamical systems analysis}\label{globdynsys}

To perform a dynamical systems analysis of the global system~\eqref{dynsys}
we need concepts and techniques from dynamical systems theory. For the
reader's convenience we therefore first recall some basic mathematical
definitions, while various techniques and results from the theory of
dynamical systems are introduced subsequently according to when they are
needed (for further details, see, e.g.,~\cite{waiell97,cra91,car81}).

Consider an autonomous dynamical system $\dot{x} = f(x)$, $x\in
\mathbb{R}^m$. The evolution of a state space point of the dynamical system
is described by the flow, which formally is a mapping $\phi_t:\mathbb{R}^m
\rightarrow \mathbb{R}^m$ that yields $x(0) \rightarrow x(t)$. The
$\omega$-limit set $\omega(x)$ of a point $x\in\mathbb{R}^m$ is defined as
the set of all accumulation points of the future orbit (i.e. solution
trajectory) through $x$. Correspondingly, the $\alpha$-limit set $\alpha(x)$
is defined as the set of accumulation points of the past orbit. The
$\omega$-limit of a set $S\subseteq \mathbb{R}^m$ is $\omega(S) =
\bigcup_{x\in S} \omega(x)$. The $\omega$-limit sets ($\alpha$-limit sets)
characterize the future (the past) asymptotic behavior of the dynamical
system. The simplest examples of limit sets are fixed points (i.e., points
$x_0$ in the state space of the dynamical system $\dot{x} = f(x)$ for which
$f(x_0)=0$; sometimes fixed points are referred to as equilibrium, critical
or stationary points) and periodic orbits.

We begin our analysis of the global dynamical system~\eqref{dynsys} by noting
that
\begin{equation}\label{mon}
\left. \frac{dT}{d\bar{\tau}}\right|_{\sin\theta = 0} = 0,\qquad
\left. \frac{d^2T}{d\bar{\tau}^2}\right|_{\sin\theta = 0} = 0,\qquad
\left. \frac{d^3T}{d\bar{\tau}^3}\right|_{\sin\theta = 0} = 6T^3(1-T)^2.
\end{equation}
As seen from~\eqref{Teq} and~\eqref{mon} $T$ is a monotonically increasing
function on ${\bf S}$ (hence $T$ can be viewed as a time variable if one is
so inclined) and as a consequence all orbits originate from the invariant
subset boundary $T=0$ (which therefore contains the $\alpha$-limit set of
${\bf S}$), which is associated with the asymptotic (classical) initial state
where $H\rightarrow \infty$, and end at the invariant subset boundary $T=1$,
which corresponds to the asymptotic future where $H =0$. Furthermore, the
equation on the invariant boundary subset $T=1$ is given by $\left.
\frac{d\theta}{d\bar{\tau}}\right|_{T=1} = -1$, and hence yields a periodic
orbit in the negative $\theta$ direction; due to that $T$ is monotonically
increasing on ${\bf S}$, it follows that this periodic orbit is a limit cycle
that constitutes the global future attractor, being the $\omega$-limit set
for all orbits in the physical state space ${\bf S}$.

Next we turn to the fixed points on $\bar{\bf S}$. Recall that a fixed point
is \emph{hyperbolic} if the linearization of the system at the fixed point is
a matrix that possesses eigenvalues with non-vanishing real parts. In this
case the \emph{Hartman-Grobman theorem} applies: In a neighborhood of a
hyperbolic fixed point the full nonlinear dynamical system and the linearized
system are topologically equivalent. The Hartman-Grobman theorem establishes
the local stability properties of a hyperbolic fixed point, but it should be
pointed out that there is no guarantee that the linearized solution yield an
approximation for the solutions in the neighborhood of the fixed point.
\emph{Non-hyperbolic fixed points} have linearizations with one or more
eigenvalues with vanishing real parts.

The only fixed points of the system~\eqref{dynsys} on the cylinder $\bar{\bf
S}$ are those on the invariant subset $T=0$, and, as can be easily seen, they
are connected by \emph{heteroclinic orbits}, and thus there are no homoclinic
orbits (an orbit whose $\omega$- and $\alpha$-limit set is a fixed point is
called a heteroclinic orbit; a homoclinic orbit originates from and ends at
one and the same fixed point). Since $\left.
\frac{d\theta}{d\bar{\tau}}\right|_{T=0} = \frac32\sin 2\theta$, the fixed
points are determined by the following values for $\theta$:
\begin{equation}
\mathrm{M}_\pm\!\!:\quad \theta = (4n \pm 1)\frac{\pi}{2}; \qquad
\mathrm{dS}_+\!\!:\quad \theta = 2n\pi, \qquad \mathrm{dS}_-\!\!:\quad \theta = (2n + 1)\pi,
\end{equation}
where $n$ is an integer. There are two equivalent (due to the discrete
symmetry) hyperbolic fixed points $\mathrm{M}_\pm$, for which $q=2$ and
$w_\phi=1$, i.e., they are associated with a massless scalar field state, and
two equivalent fixed points $\mathrm{dS}_\pm$, for which $q=-1$ and
$w_\phi=-1$, which therefore correspond to a de Sitter state. It should,
however, be noted that the present de Sitter fixed points are distinct from
de Sitter states that are associated with potentials that admit situations
for which $dV/d\phi=0$ for some constant finite value of $\phi$ for which
both $V$ and $H$ have constant, bounded, and positive values. In contrast the
present de Sitter states correspond to the limits $\dot{\phi} =0$,
$(\phi,V,H) \rightarrow (\pm\infty,\infty,\infty)$. The massless scalar field
fixed points, $\mathrm{M}_\pm$, are hyperbolic sources, while the de Sitter
fixed points, $\mathrm{dS}_\pm$, are sinks \emph{on} $T=0$ (they have one
negative eigenvalue associated with the $T=0$ subset given by $-3$), but they
also have one zero eigenvalue that corresponds to a 1-dimensional unstable
center submanifold (and thus they are non-hyperbolic in $\bar{\bf S}$), to be
discussed below, and thus one solution originates from each fixed point
$\mathrm{dS}_\pm$ into ${\bf S}$.

The solutions that originate from $\mathrm{dS}_\pm$ are often referred to as
`attractor solutions,' a nomenclature that for the present scalar field
potential originates from the work by Belinski\v{\i} {\it et
al}.~\cite{beletal85a,beletal85b,belkha87}. The nomenclature has been
motivated by heuristic linear analysis, which indicates that the attractor
solutions are locally stable. As can be shown numerically, it is even true
that solutions that correspond to initial data quite far from the attractor
solution, where a linear analysis would not be valid, rapidly approach it in
the present variables. Nevertheless, it is also clear that there exists an
open set of non-inflating solutions that only come close to the attractor
solution near the future global attractor at $T=1$. The fact that there
exists such an open set of solutions brings up the issue of imposing a
measure on the state space, a problem already recognized by Belinski\v{\i}
{\it et al}.~\cite{beletal85a,beletal85b,belkha87,beletal88}. However, it
should be pointed out that it is possible to make a nonlinear variable change
so that the attractor solutions are not locally stable everywhere and hence
the nomenclature `attractor solution' is of rather heuristic nature (for
further discussions about the meaning of `attractor solutions' and measures,
see the recent papers by Remmen and Carroll~\cite{remcar13,remcar14} and by
Corichi and Sloan~\cite{corslo14}, and references therein).

Furthermore, it is worth noting that there already exists a non-trivial
example of this. In the global state space for the modified Chaplygin gas
used in~\cite{ugg13} it was shown that the perfect fluid attractor solution
is not attracting nearby solutions everywhere along its evolution.
Nevertheless, there exists an open set of solutions that are always
arbitrarily close to it during their intermediate and late time evolution.
These results are to be contrasted with previous work, which had established
linear stability of the attractor solution in terms of other variables. It
should also be pointed out that (apart from the case of an exponential
potential) scalar field problems have reduced state spaces with more than one
dimension, and that usually the linear stability analysis that motivates
calling a solution an attractor solution in the literature only makes use of
a single variable, e.g. $H$, but stability in one variable such as $H$ does
not guarantee stability in additional dimensions.

\begin{figure}[ht!]
\centering
{\includegraphics[width=0.60\textwidth]{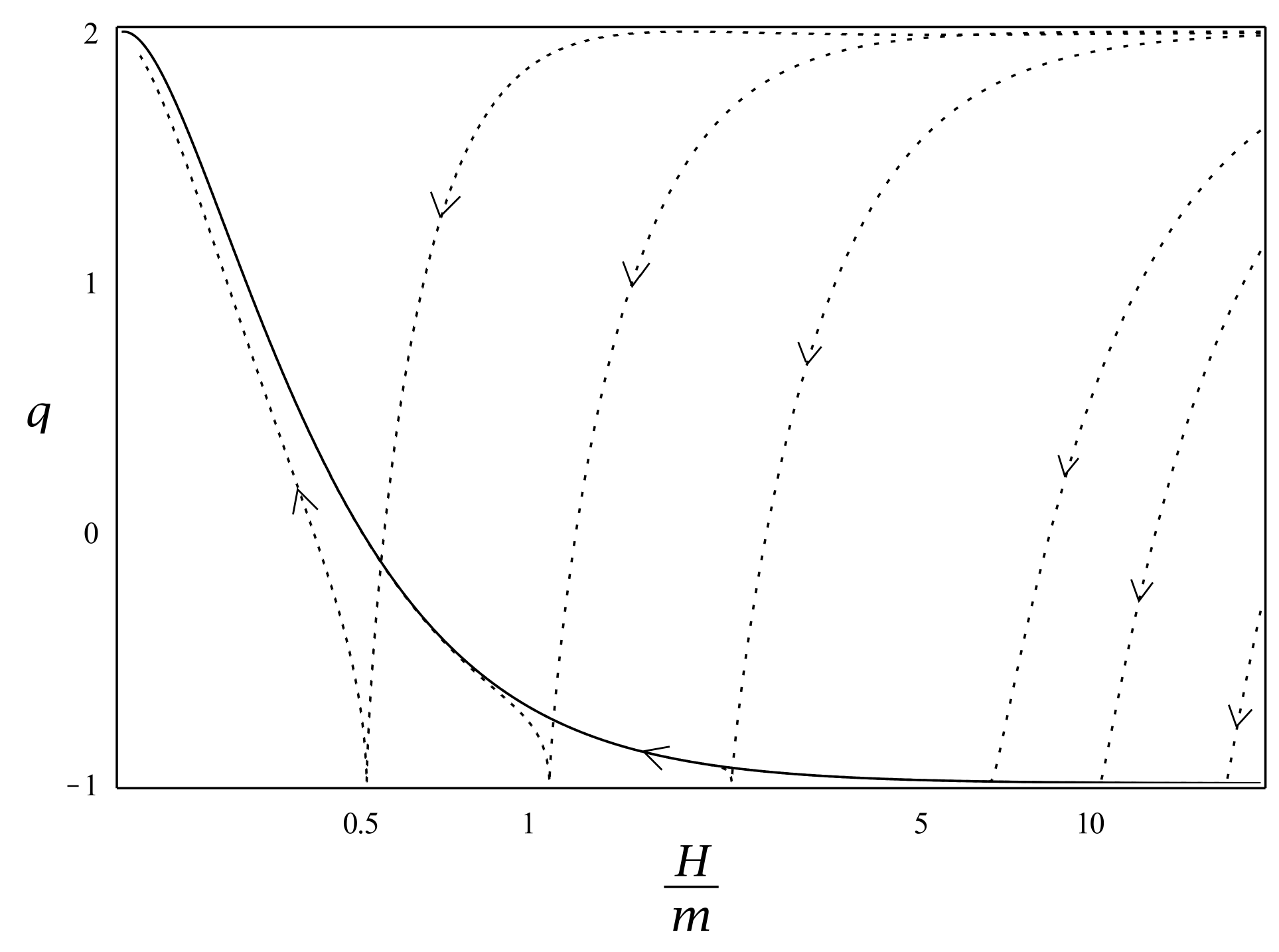}}
\caption{$H/m-q$ diagram, where $H$ is the Hubble variable and $q$ the deceleration parameter.
It is seen that other solutions (dotted lines) than the attractor solution (the solid line)
evolve toward the attractor solution, where the arrows denote the time direction
(recall that $H/m$ is monotonically decreasing in time). The computations have been interrupted
at $q=2$, since the solutions then enter the oscillatory phase.
}\label{fig:Hq}
\end{figure}

Instead of using measures to make sense of the attracting properties of the
attractor solution one can consider the evolution of physical
measurable observables and study if they evolve toward the attractor
solution. In Figure~\ref{fig:Hq} we have plotted the attractor solution in
$H/m-q$-space just before the oscillatory regime together with other
solutions for various initial data, and, as can be seen, solutions are indeed
`attracted' toward the attractor solution. Nevertheless, it should be pointed
out that the formal mathematical future global attractor is the periodic
orbit on the invariant $T=1$ boundary subset. (Loosely speaking, in dynamical
systems theory attractor behavior describes situations where a collection of,
in some sense, generic state space points evolve into a certain `attractor'
region from which they never leave. For a formal definition of a dynamical
systems attractor, see e.g.~\cite{waiell97}, and references therein.)

Finally, note that ${\bf S} \rightarrow (H/m,q)$ is a surjective two-to-one
map, i.e., (as can be expected) each solution only appears once in the state
space of the geometric observables $(H/m,q)$ instead of twice in ${\bf S}$.
It is possible to use e.g. $(T,q),\, q\in [-1,2]$ (or $(T,w_\phi),\, w_\phi
\in [-1,1]$) as a bounded state space, but $q=-1$ and $q=2$ ($w_\phi=\pm 1$)
are not invariant boundaries in such a formulation, which lead to analytic
difficulties. As a consequence it is therefore preferable to use the state
space ${\bf S}$, since it is bounded by invariant boundaries in an analytic
manner which makes it possible to extend the state space to $\bar{\bf S}$.

\begin{figure}
\centering
{\includegraphics[width=0.40\textwidth]{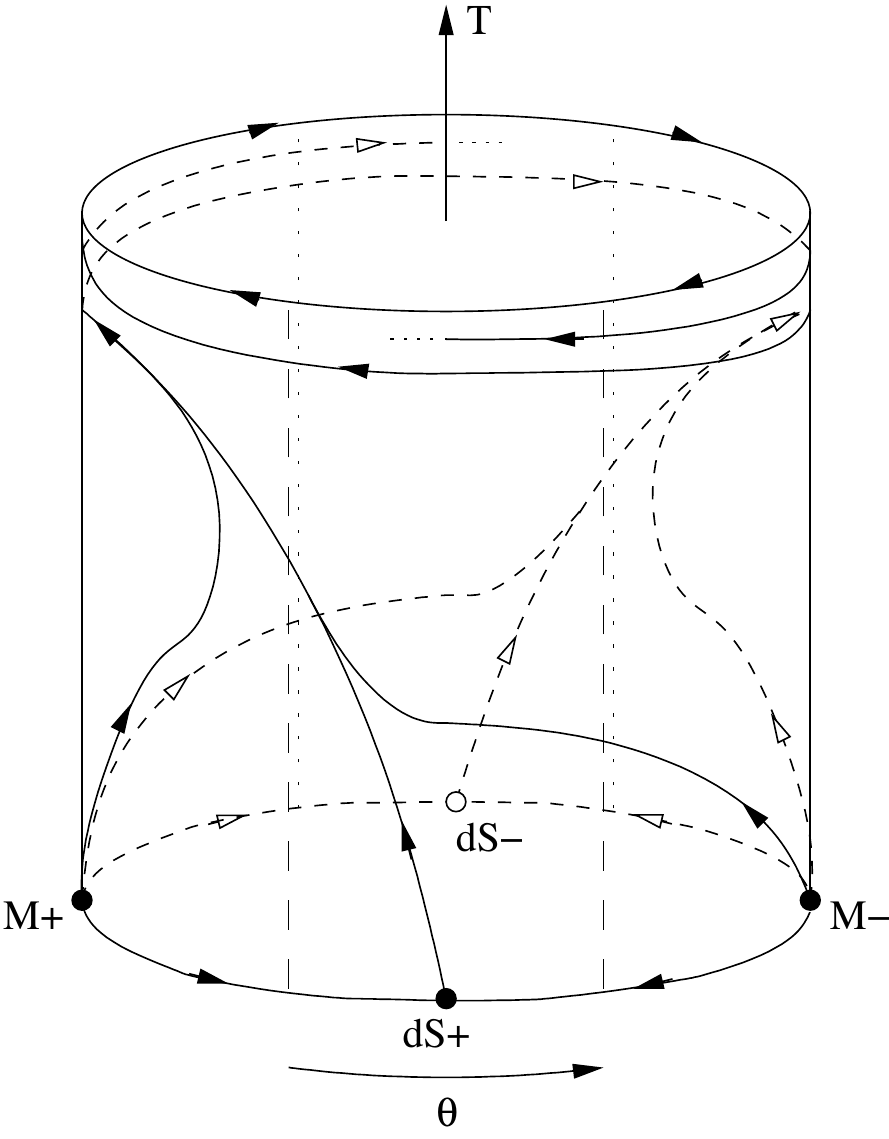}}
\caption{Examples of solutions in the global state space $\bar{\bf S}$. All
solutions originate from $\mathrm{M}_\pm$, except for the two equivalent `attractor
solutions' that originate from the center saddles $\mathrm{dS}_\pm$. All
solutions end at the limit cycle at the $T=1$ boundary at the top of the
cylinder, which hence is the future global attractor. Note the strips of
accelerating regions characterized by $\sin^2\theta < \frac13$.}\label{fig:phasespace}
\end{figure}

To summarize the global picture of the solution space on ${\bf S}$: All
orbits except for the two equivalent attractor solutions, which originate
from $\mathrm{dS}_+$ and $\mathrm{dS}_-$, originate from the two equivalent
sources $\mathrm{M}_+$ and $\mathrm{M}_-$, while all solutions asymptotically
approach the periodic orbit on the $T=1$ invariant boundary subset, which
hence is the future global attractor. This situation is depicted in
Figure~\ref{fig:phasespace}, which gives a global state space description of
the solution space. Note that the heuristic description in terms of a damped
harmonic oscillator does not reveal that there are two equivalent attractor
solutions that originate from a de Sitter state while all other solutions
originate from a massless scalar field state, nor does it reveal that the
final state is one for which dimensionless physical observables such as the
deceleration parameter $q$ oscillate with a constant amplitude.
%

\subsection{Approximations at late times}

\begin{figure}[ht!]
\begin{center}
{\subfigure[The averaged solution at late times.]{
        \label{fig:average}
        \includegraphics[width=0.45\textwidth]{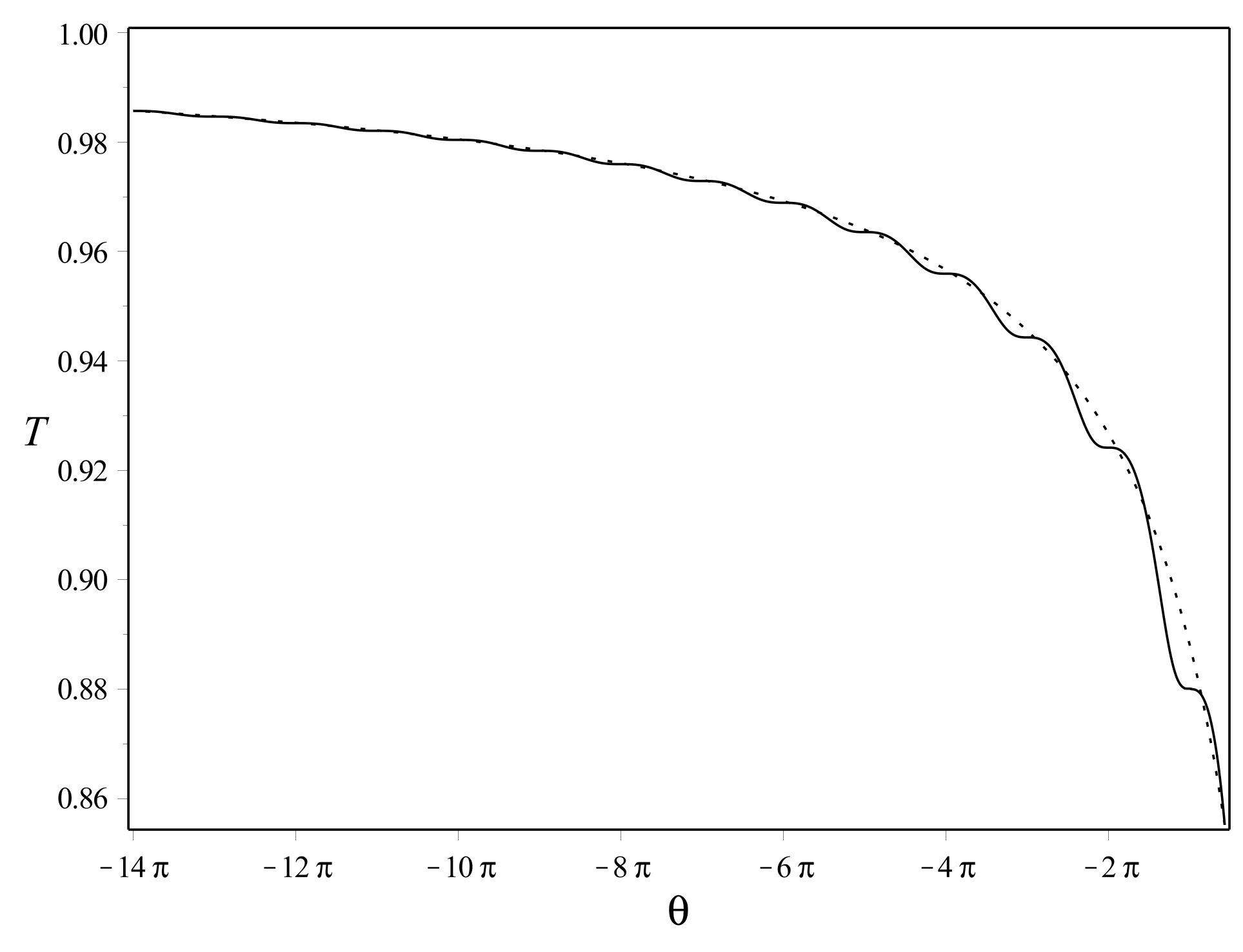}}}
{\subfigure[An oscillatory approximation for the oscillatory late time regime.]{
        \label{fig:osc}
        \includegraphics[width=0.45\textwidth]{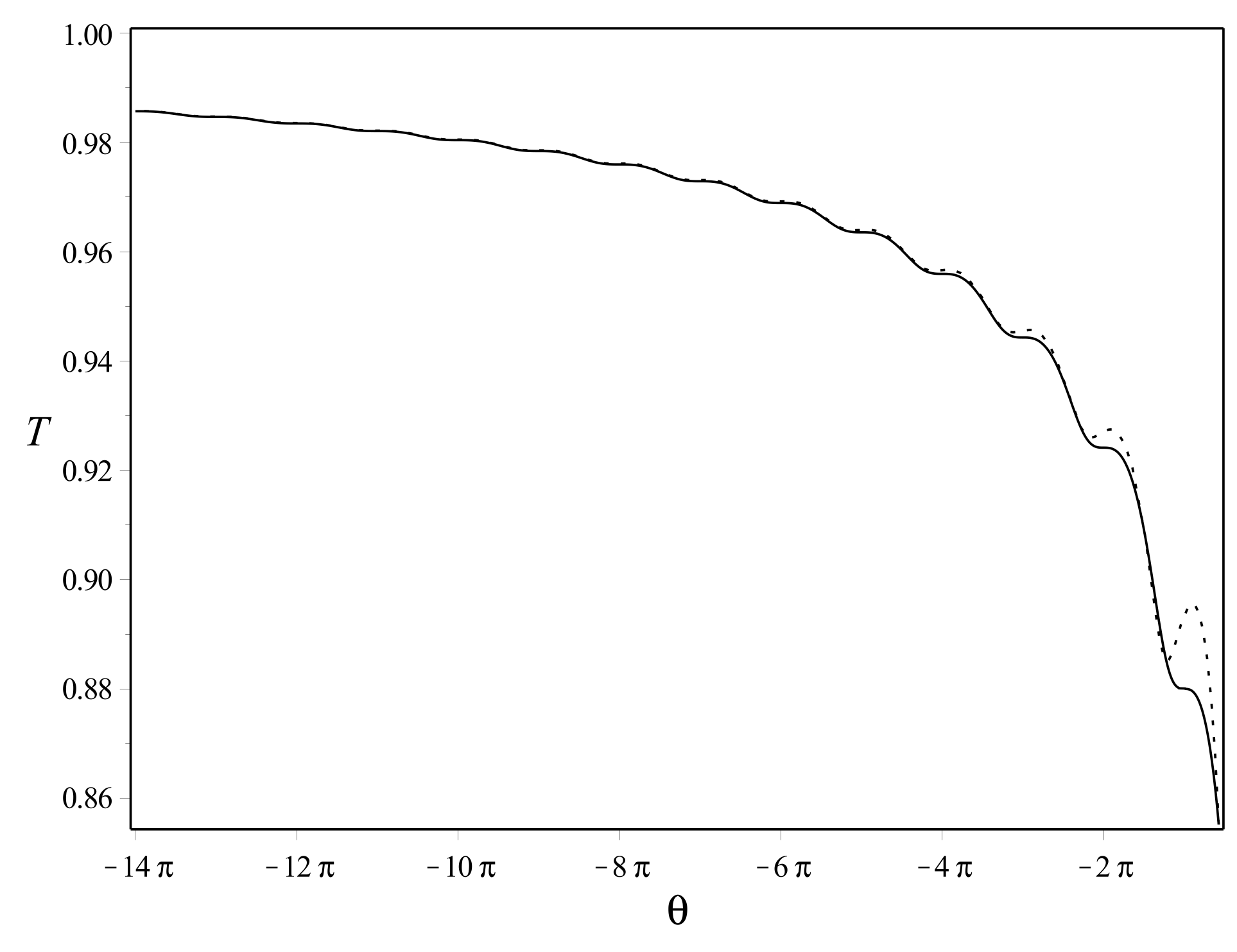}}}
\end{center}
        \caption{The plot~\ref{fig:average} shows the numerical attractor solution (solid line) and
the averaged solution (dotted line). The plot~\ref{fig:osc} shows the numerical attractor solution (solid line) and
the oscillatory late time approximation (dotted line).}
    \label{fig:latetimes}
\end{figure}

To obtain approximations for solutions at late times close to $T=1$ (the
oscillatory phase), we take the average with respect to $\theta$ of the right
hand side of~\eqref{dynsys} (since $-\theta \rightarrow \bar{\tau} \propto t
\rightarrow \infty$ while $T$ slowly approaches one), which leads to
\begin{subequations}\label{dynsysaverage}
\begin{align}
\frac{dT}{d\bar{\tau}} &= \frac32 T(1-T)^2,\label{Teqav}\\
\frac{d\theta}{d\bar{\tau}} &= - T  .
\end{align}
\end{subequations}
It follows that
\begin{equation}\label{thetaTlate}
\theta = - \frac{2}{3(1-T)} + C,
\end{equation}
where $C = \theta_i + 2/3(1-T_i)$, where $(\theta_i, T_i)$ is some initial
point for the trajectory (for heuristic work on oscillatory late time
behavior for monomial scalar field potentials, see
e.g.~\cite{tur83,dammuk98,macpic00}). This approximation is valid for all
solutions when $T$ approaches one, including the center manifold 
attractor solution, see Figure~\ref{fig:average}.
In~\cite{ren07} Rendall gave rigorous results for late time behavior, which
to leading order corroborates the relation given in eq.~\eqref{thetaTlate},
but Rendall also refined this approximation. In particular he proved results
that can be translated to the following asymptotic
approximation~\cite{ren07}:
\begin{equation}\label{thetaTlateosc}
\theta = -t - \frac{3+2\cos(2t)}{4t}, \qquad T = \left(1 + \frac{2}{3(t-t_0)}\left(1 + \frac{\sin(2t)}{2t}\right)\right)^{-1},
\end{equation}
where $t_0$ is a constant and $t$ is synchronous time (with $m$ normalized to
one); note that~\eqref{thetaTlate} is obtained in the limit $t\rightarrow
\infty$. The relations given in eq.~\eqref{thetaTlateosc} describe a
parameterized curve in the global state space ${\bf S}$, which is plotted in
Figure~\ref{fig:osc}; note that the oscillatory approximation becomes
increasingly good toward the future, reflecting that it describes the
asymptotic evolution at late times.

Alternatively, improved late time approximations (explicitly including
oscillatory behavior) can be obtained by using the averaged solution as a
starting point for more accurate approximations in the manner illustrated
in~\cite{waietal99} for the case of perfect fluid dynamics in Bianchi type
VII$_0$ at late times. These models, like the present ones, provide an
example of \emph{asymptotic manifest self-similarity breaking} at late times,
where asymptotic (continuous) manifest self-similarity is defined by the
requirement that all physical geometrical scale-invariant observables such as
the deceleration parameter $q$ take asymptotic constant values (the issue of
asymptotic manifest self-similarity breaking turns out to be quite subtle and
we will return to it more thoroughly in a forthcoming paper). Since the
future attractor in the present case is a limit cycle it follows that $q$ is
asymptotically oscillating, and thus that asymptotic manifest self-similarity
is broken.

Incidentally, the late time behavior in $(\phi,\dot{\phi})$-space is
associated with a non-hyperbolic fixed point at the origin which acts as an
attracting focus. This is rather misleading since the existence of an
attracting fixed point might lead one to believe that the late time dynamics
is asymptotically manifestly self-similar; the true situation, i.e.
asymptotic manifest self-similarity breaking, is only revealed by resolving
the non-hyperbolicity of the fixed point, which leads to the present picture
with a limit cycle.

\subsection{Center manifold analysis of the attractor solution}\label{sec:centexp}

Next we will extend the center manifold analysis of~\cite{ren02} (a paper
that was inspired by~\cite{fos98}), which describes the (equivalent)
attractor solutions that originate from $\mathrm{dS}_\pm$. To do so we first
briefly review some aspects of center manifold analysis that are needed for
the present problem, which can be regarded as a specific application (for
more details concerning center manifolds, see e.g.~\cite{cra91,car81}; for
recent uses of center manifold analysis in cosmology, see
e.g.~\cite{sinetal03,bohetal12}).

Assume that there exists a fixed point $P$ of a dynamical system $\dot{x} =
f(x)$, $x\in \mathbb{R}^m$ and that the linearization of the system at $P$ is
described by a matrix that can be diagonalized. Accordingly, the phase space
$\mathbb{R}^m$ of the linearized equations can be decomposed into the direct
sum $E^s\oplus E^u\oplus E^c$ of the stable subspace $E^s$ (spanned by the
eigenvectors of the eigenvalues with negative real parts), the unstable
subspace $E^u$ (spanned by the eigenvectors of the eigenvalues with positive
real parts) and the center subspace $E^c$ (spanned by the eigenvectors of the
eigenvalues with vanishing real parts). These subspaces of the linearized
system form the tangent spaces to associated invariant submanifolds of the
full system (which therefore have the same dimension as the tangent spaces),
which are denoted by $W^s$ for the stable manifold, $W^u$ for the unstable
manifold, and $W^c$ for the center manifold.

Without loss of generality, we locate the fixed point $P$ at
$(x_1,x_2)=(0,0)$ and choose variables such that the dynamical system can be
written as
\begin{subequations}\label{fullnonlinear}
\begin{align}
\dot{x}_1 & = A_1 x_1 + N_1(x_1,x_2), \\
\dot{x}_2 & = A_2 x_2 + N_2(x_1,x_2).
\end{align}
\end{subequations}
Here $x_1 \in E^c$ is $m_c$-dimensional while $x_2\in E^s\oplus E^u$ is
($m_u+m_s$)-dimensional; $A_1$ is a constant $m_c\times m_c$ matrix with
purely imaginary (or zero) eigenvalues; $A_2$ is a constant $(m_s+m_u) \times
(m_s+m_u) $ matrix for which all eigenvalues have nonzero real parts;
finally, $N_1$ and $N_2$ denote the nonlinear terms.

Since the center manifold $W^c$ is an invariant manifold in the phase space
$\mathbb{R}^m$ that contains $P$ and that is tangent to $E^c$, one can
describe $W^c$, in a neighborhood of $P$, as the graph of a function $h: E^c
\rightarrow E^s\oplus E^u$, i.e., $x_2=h(x_1)$, where for sufficiently small
$x_1$ the point $(x_1, h(x_1))$ belongs to $W^c$. Inserting the relation
$x_2=h(x_1)$ into~\eqref{fullnonlinear} yields the following differential
equation for $h$:
\begin{equation}\label{hequation}
\partial_{x_1} h(x_1) \:[ A_1 x_1 + N_1(x_1,h(x_1))]=
A_2 h(x_1) + N_2(x_1,h(x_1)),
\end{equation}
which also satisfies $h(0)=0$ (fixed point condition) and $(\partial_{x_1}
h)(0) = 0$ (tangency condition). Usually, however, it is not possible to
solve~\eqref{hequation}. Instead this equation can be solved approximately by
representing $h(x_1)$ as a formal (usually truncated) power series, and solve
for the constant coefficients to any desired order.

The Hartman-Grobman theorem can be generalized to the Shoshitaishvili theorem
(sometimes referred to as the reduction theorem) for the present situation of
a non-hyperbolic fixed point:
\begin{theorem}\label{shoshitaishvili}(Shoshitaishvili theorem).
Consider the dynamical system~(\ref{fullnonlinear}) in a neighborhood of the
fixed point $P=(0,0)$. The flow of the full nonlinear system and the flow of
the reduced system
\begin{subequations}\label{reduced}
\begin{align}
\dot{x}_1 & = A_1 x_1 + N_1(x_1,h(x_1)) \\
\dot{x}_2 & = A_2 x_2 \:.
\end{align}
\end{subequations}
are locally equivalent, i.e., there exists a local homeomorphism $\Psi$ on
phase space, such that $\phi_t^{\mathrm{full}} = \Psi^{-1} \circ
\phi_t^{\mathrm{reduced}} \circ \Psi$. Here, $h$ is given
by~(\ref{hequation}).
\end{theorem}
Note that $x_2=0$ forms an invariant subset of~\eqref{reduced} that describes
the center manifold.

\begin{figure}[ht!]
\begin{center}
\subfigure[Center manifold expansions.]{\label{fig:numCM}
\includegraphics[width=0.45\textwidth]{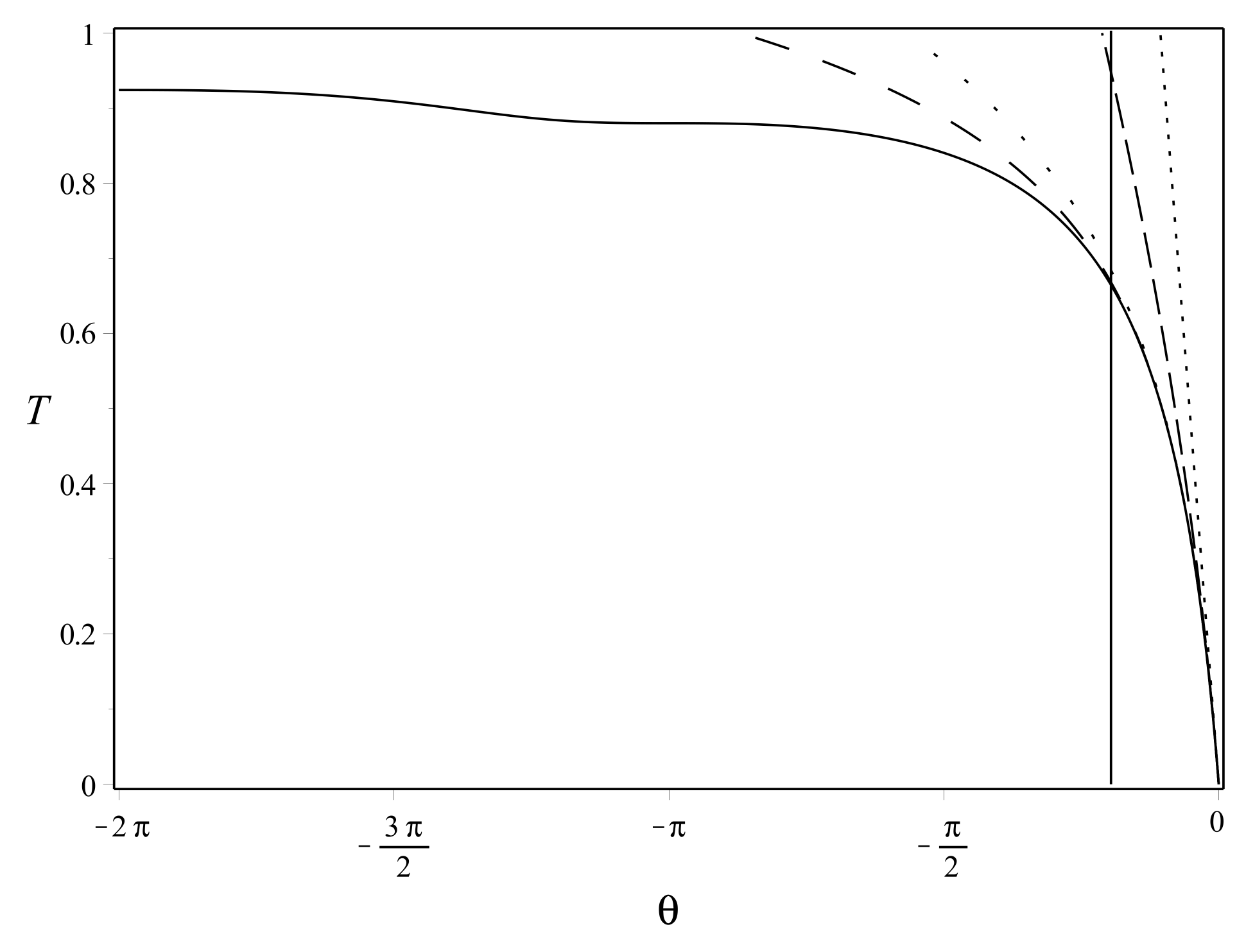}
}%
\subfigure[Relative errors for center manifold expansions.]{\label{fig:CMerror}
\includegraphics[width=0.45\textwidth]{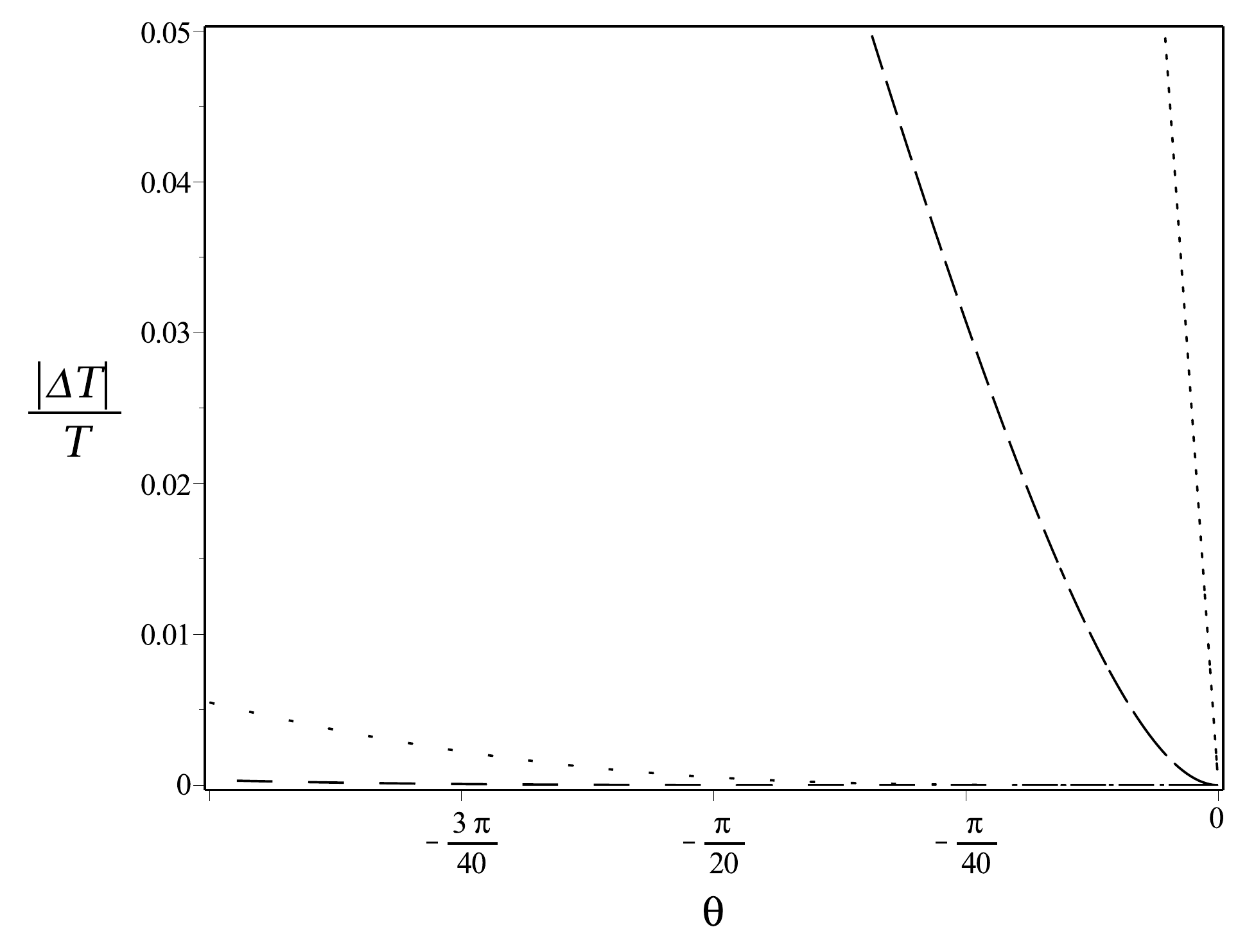}
}
\end{center}
\caption{The plot~\ref{fig:numCM} shows the state space $(T,\theta)$, with $\theta$ in the range $[-2\pi,0]$,
with the numerical solution (solid line) and the center manifold expansion ($\mathrm{CMexp}$)
to 1st (dotted), 2nd (dashed), 6th (space-dotted), and 10th (space-dashed) order approximation.
The end of the accelerating regime, $q=0$, is depicted by the vertical line. The plot~\ref{fig:CMerror}
shows the relative errors $|\Delta T|/T = |T_\mathrm{CMexp} - T_\mathrm{num}|/T_\mathrm{num}$
for the center manifold expansions for small $\theta$ (and hence for small $T$).}%
\label{fig:numcent}
\end{figure}

Let us now apply the above to the global dynamical system~\eqref{dynsys}. Due
to the discrete symmetry it suffices to investigate one of the non-hyperbolic
$\mathrm{dS}_\pm$ fixed points, and without loss of generality it is
convenient to choose the $\mathrm{dS}_+$ fixed point at $(T,\theta) = (0,0)$.
The starting point for a center manifold analysis is a linearization of the
system at the fixed point, which in the present case leads to the eigenvalues
$-3$ and zero and the following stable, $E^s$, and center, $E^c$, tangential
subspaces, respectively:
\begin{subequations}\label{linearanalysis}
\begin{align}
E^s &= \{(T,\theta)|T=0\},\\
E^c &= \{(T,\theta)|T + 3\theta =0\}.
\end{align}
\end{subequations}
Since the tangential center subspace is given by $T + 3\theta =0$, we
introduce $v=T + 3\theta$ as a new variable in order to study the center
manifold $W^c$ (with the tangent space $E^c$ at $(T,v) = (0,0)$). As follows
from~\eqref{dynsys}, this leads to the transformed system
\begin{subequations}\label{dynsysv}
\begin{align}
\frac{dT}{d\bar{\tau}} &= 3T(1-T)^2\sin^2\left(\sfrac13(v-T)\right),\label{Tveq}\\
\frac{dv}{d\bar{\tau}} &= 3\left(T(1-T)^2\sin^2\left(\sfrac13(v-T)\right) - T - \frac32(1-T)\sin \left(\sfrac23(v-T)\right)\right) .\label{vveq}
\end{align}
\end{subequations}
Note that the linearization of eq.~\eqref{vveq} yields
$\frac{dv}{d\bar{\tau}} = - 3v$, while eq.~\eqref{Tveq} only has higher order
terms, which corresponds to the eigenvalues $-3$ and zero. The center
manifold $W^c$ can be obtained as the graph $v = h(T)$ near $(T,v) = (0,0)$
(i.e., use $T$ as an independent variable), where $h(0) =0$ (fixed point
condition) and $\frac{dh}{dT}(0) =0$ (tangency condition). Inserting this
relationship into eq.~\eqref{dynsysv} and using $T$ as the independent
variable leads to
\begin{equation}\label{hT}
T(1-T)^2\sin^2\left(\sfrac13(h-T)\right)\left(\frac{dh}{dT}-1\right) + T + \frac32(1-T)\sin\left(\sfrac23(h-T)\right)  = 0.
\end{equation}
Not surprisingly it is hard to solve this differential equation since this
would amount to actually finding the center manifold, i.e., in this case the
attractor solution. However, we can solve the equation approximately by
representing $h(T)$ as the formal power series
\begin{equation}
h(T) = \sum_{i=2}^n a_iT^i + {\cal O}(T^{n+1}) \qquad \text{as}\qquad T\rightarrow 0,
\end{equation}
where the series for $h(T)$ is truncated at some chosen order $n$. Inserting
this into eq.~\eqref{hT} and algebraically solving for the coefficients leads
to that $\theta = \frac13(-T + h(T))$ is given by
\begin{equation}\label{CM_Expansion}
\begin{split}
\theta &= -\frac13\left\{T + T^2 + \underbrace{\frac{26}{3^3}}_{0.963}T^3 + \underbrace{\frac{8}{3^2}}_{0.889}T^4 +
\underbrace{\frac{107}{3^3\cdot 5}}_{0.793}T^5 + \underbrace{\frac{19}{3^3}}_{0.704}T^6\right. \\ & \\
& \qquad \left. + \underbrace{\frac{3352}{3^6\cdot 7}}_{0.657}T^7 +
\underbrace{\frac{490}{3^6}}_{0.672}T^8 + \underbrace{\frac{43381}{3^{10}}}_{0.735}T^9 +
\underbrace{\frac{25961}{3^8\cdot 5}}_{0.791}T^{10}\right\} + {\cal O}(T^{11}),
\end{split}
\end{equation}
where we have chosen to expand $h(T)$ to 10th order. Figure~\ref{fig:numcent}
depicts the numerical description of the attractor solution and plots of
curves associated with expansions to various orders obtained
from~\eqref{CM_Expansion} (to avoid clutter we do not give all expansions up
to 10th order, but each order gives a more accurate approximation than the
previous one). As is seen from this figure, and Table~\ref{tab:cexpq0},
higher order expansions yield increasingly better approximations for the
attractor solution, even far from $(T,\theta) = (0,0)$, indicating quite good
convergence for rather large $T$, i.e., small $H/m$, even beyond the end of
inflation, i.e., beyond $q=0$ (although it is only at second order or higher
that $q=0$ is actually passed).

\begin{table}[ht!]
\begin{center}
 \begin{tabular}{|c|c|c|c|c|c|}  \hline
                       &             &             &           &          &          \\ [-2ex]
                       &  Num.       & 1st         &    2nd    &    6th   & 10th     \\  [1ex]\hline
                       &             &             &           &          &          \\ [-2ex]
$T$                    & 0.6646      & \textemdash & 0.9479    & 0.6842   & 0.6682   \\[1ex] \hline
                       &             &             &           &          &          \\ [-2ex]
$\frac{|\Delta T|}{T}$ & \textemdash & \textemdash & 42.627\%  & 2.949\%  & 0.542\%  \\ [1ex]\hline
                       &             &             &           &          &          \\ [-2ex]
$\frac{H}{m}$          & 0.5047      & \textemdash & 0.0550    & 0.4616   & 0.4967   \\[1ex] \hline
                       &             &             &           &          &          \\ [-2ex]
$\frac{|\Delta H|}{H}$ & \textemdash & \textemdash &  89.112\% & 8.538\%  & 1.595\%  \\ [1ex]\hline
 \end{tabular}
\end{center}
\caption{Numerical values and relative errors for center manifold expansions
at $q=0$. The relative errors are given by $|\Delta T|/T = |T_\mathrm{CMexp}
- T_\mathrm{num}|/T_\mathrm{num}$ and $|\Delta H|/H = |H_\mathrm{CMexp}
- H_\mathrm{num}|/H_\mathrm{num}$
.}\label{tab:cexpq0}
\end{table}
%

\subsection{Pad{\'e} approximants for the attractor solution}

To obtain a better range and rate of convergence we will replace the above
center manifold expansions in eq.~\eqref{CM_Expansion} with so-called
Pad{\'e} approximants (for more details, see
e.g.~\cite{sinetal03,bak75,kal02}, and for some recent examples in cosmology,
see~\cite{gruluo14,weietal14}). A Pad{\'e} approximant is a particular
approximation of a function at a regular point by a rational function. The
Pad{\'e} approximant of \emph{order} $(m,n)$ of a function $f(x)$, denoted by
$[m/n]_f(x)$, is associated with a truncated Taylor series (where we without
loss of generality assume that the regular point we are interested in is
located at $x=0$, i.e., we perform a truncated Maclaurin series expansion):
\begin{equation}
f \approx c_0 + c_1 x + c_2 x^2 + \cdots + c_{m+n}x^{m+n},
\end{equation}
and given by the polynomials $P_m(x)$ and $Q_n(x)$ according to
\begin{equation}
[m/n]_f(x) = \frac{P_m(x)}{Q_n(x)} = \frac{p_0 + p_1 x + p_2 x^2 + \cdots+p_m x^m}{q_0 + q_1 x + q_2 x^2+\cdots+q_n x^n},
\end{equation}
such that
\begin{equation}
Q_n(x)(c_0 + c_1 x + c_2 x^2 + \cdots + c_{m+n}x^{m+n}) = P_m(x),
\end{equation}
where coefficients with the same powers of $x$ are equated up through $m+n$.
Thus
\begin{subequations}
\begin{alignat}{2}
&\sum_{j=0}^n q_j c_{m-j+k} = 0, &\qquad  k&=1,\dots,n,\\
&\sum_{j=0}^k q_j c_{k-j} - p_k = 0, &\qquad  k&=0\dots,m,
\end{alignat}
\end{subequations}
yields a system of $m+n+1$ linear equations in $m+n+2$ unknowns
$p_0,\dots,p_m$ and $q_0,\dots,q_n$ (writing the above linear system, which
defines the Pad{\'e} approximant, as a matrix equation reveals that the
defining matrix takes a very special form, namely that of a T{\"o}plitz
matrix). This therefore leads to several possible solutions, but the possible
expressions for $[m/n]_f(x)$ only differ by a common factor in $P_m(x)$ and
$Q_m(x)$ and one can therefore without loss of generality scale these
polynomials to set $q_0=1$ (the so-called Baker condition; for a discussion
concerning this normalization, see~\cite{kal02}), which yields the standard
form for the Pad{\'e} approximant $[m/n]_f(x)$. This, however, assumes that
$Q_n(0)\neq 0$, while problems arise if $Q_n(0)=0$. In this paper we will
avoid such difficulties and compute Pad{\'e} approximants with $q_0=1$. It
should also be pointed out that there are more efficient ways of computing
Pad{\'e} approximants than the just outlined one, although for our purposes
the above suffices.

Why are Pad{\'e} approximants interesting? By definition it follows that
\begin{equation}
Q_n(x)f(x) - P_m(x) = {\cal O}(x^{m+n+1}),
\end{equation}
Pad{\'e} approximants therefore agree with the truncated power series of the
function it approximates up to the power $m+n$ (a Pad{\'e} approximant can be
viewed as a generalization of a Taylor polynomial which can be associated
with the above equation by formally setting $Q_n(x)=1$). Nevertheless, the
difference between the Taylor series and the associated Pad{\'e} approximants
is crucial, since the latter often works better than the power series, i.e.,
Pad{\'e} approximants often have a better rate and range of convergence. This
is illustrated by the present paper; below we will use Pad{\'e} approximants
for the center manifold expansion associated with $\mathrm{dS}_+$, and, as
will be seen, the Pad{\'e} approximants give significantly better results
than their associated Taylor series approximations. In this context it should
be pointed out that although there exists a number of convergence results in
the literature, these are not suitable for determining the precise
quantitative convergence properties of a given function, or the differential
equation that defines it, which in the present case is eq.~\eqref{hT} for
$h(T)$ (at least the authors have not found any theorem that is directly
applicable in the present context). It is for this reason we will give our
results below in the form of plotted errors. Finally, note that the
generalization of a Pad{\'e} approximant to an approximant in several
variables is called a Canterbury approximant.

Instead of producing the Pad{\'e} approximants associated with the
expansion~\eqref{CM_Expansion}, we will give the Pad{\'e} approximants that
arise from the analogous expression of the dynamical system~\eqref{dynsys2}
in Section~\ref{sec:appendixA} which is adapted to the particular structures
exhibited at early times. The reason for this is that this will yield the odd
Pad{\'e} approximants obtained from the $T$ expansion~\eqref{CM_Expansion},
but with less computation and in a succinct form. Furthermore, the even
Pad{\'e} expressions, $[2/2]_{\theta}$, $[4/4]_{\theta}$, obtained from
eq.~\eqref{CM_Expansion}, do not converge for large $T$ (in contrast to the
center manifold series expansions, which do converge for large $T$), and we
therefore refrain from giving them. To obtain the relevant Pad{\'e}
approximants we first have to calculate the center manifold expansion of
$\mathrm{dS}_+$ of the dynamical system~\eqref{dynsys2} for the dependent
variables $\theta$ and $\tilde{T}$, where we recall that ${\tilde{T}}$ is
defined by
\begin{equation}
\tilde{T} = \frac{T}{1-T} = \frac{m}{H}.
\end{equation}

\begin{figure}
\begin{center}
{\subfigure[Center manifold expansions in $\tilde{T}$]{\label{fig:tildetexp}
\includegraphics[width=0.45\textwidth]{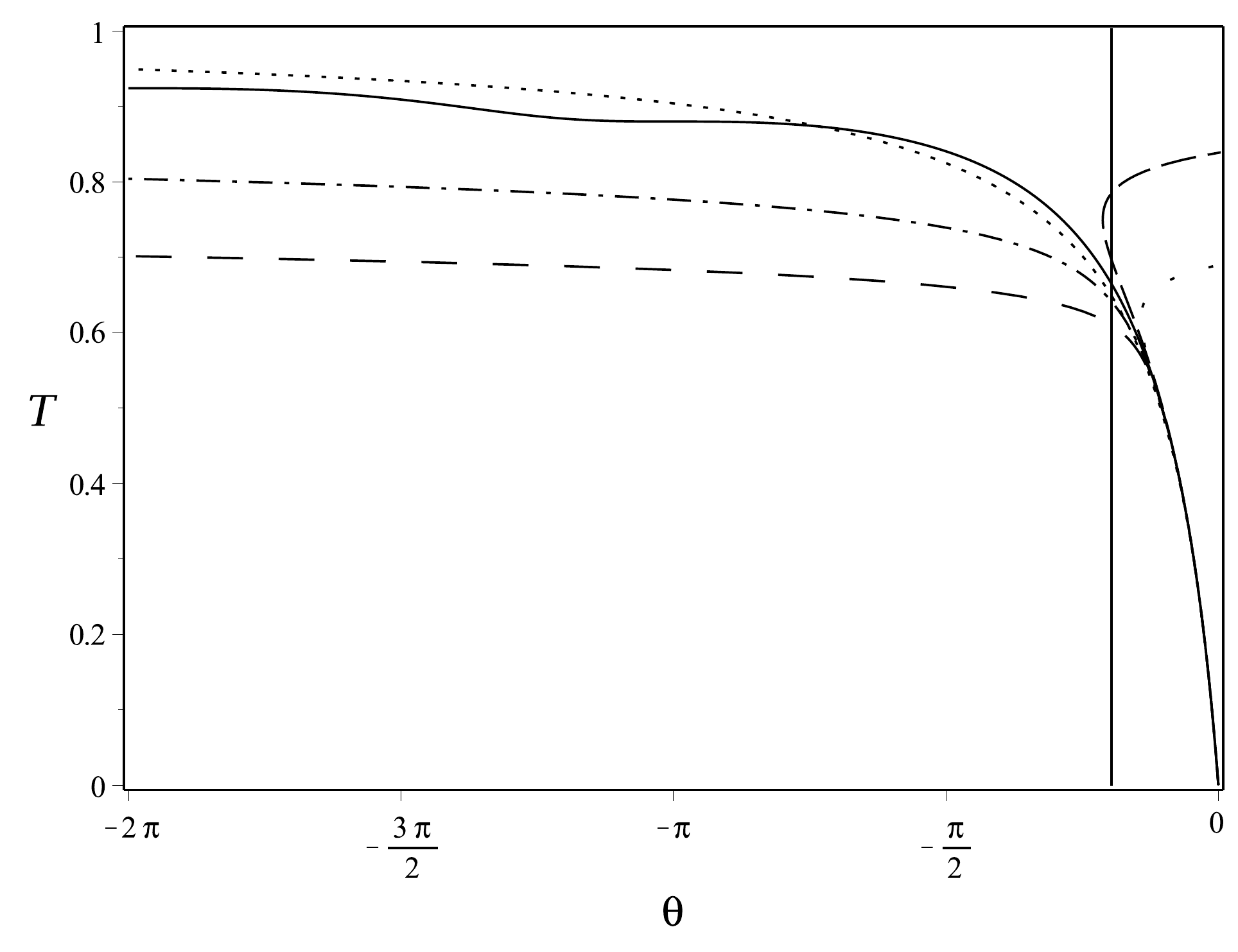}}}
{\subfigure[Relative errors for the center manifold expansions in $\tilde{T}$]{\label{fig:tildeTerrors}
\includegraphics[width=0.45\textwidth]{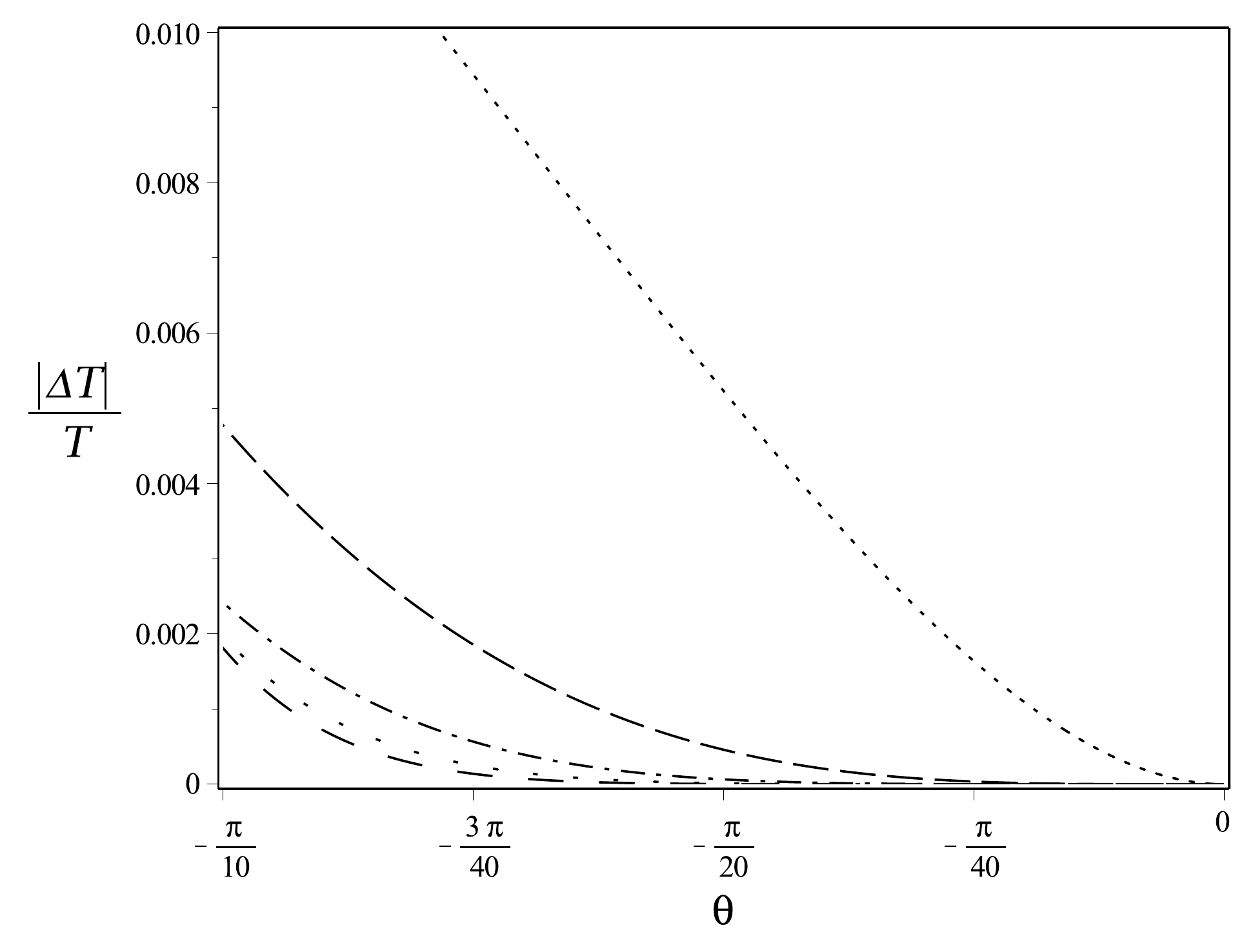}}}
\end{center}
\caption{The plot~\ref{fig:tildetexp} shows the state space $T,\theta$, with $\theta$ in the range $[-2\pi,0]$,
with the numerical solution (solid line) and the center manifold expansion in $\tilde{T}$
to 1st (dotted), 2nd (dashed), 3rd (dash-dotted), 7th (space-dotted), and 10th (space-dashed) order approximation.
The end of inflation, $q=0$, is depicted by the vertical line. The plot~\ref{fig:tildeTerrors}
shows the relative errors $|\Delta T|/T$
for the center manifold expansions for small $\theta$.}
\label{fig:tildeT}
\end{figure}

In this case we obtain
\begin{subequations}\label{linearanalysis2}
\begin{align}
E^s &= \{(\tilde{T},\theta)|\tilde{T}=0\},\\
E^c &= \{(\tilde{T},\theta)|\tilde{T} + 3\theta =0\}.
\end{align}
\end{subequations}
The graph representation of the center manifold $W^c$ is given by $\tilde{T}
+ 3\theta = \tilde{h}(\tilde{T})$ near $(\tilde{T},\theta) = (0,0)$, where,
as follows from eq.~\eqref{dynsys2}, $\tilde{h}(\tilde{T})$ obeys the first
order ordinary differential equation
\begin{equation}\label{gT}
\tilde{T}\sin^2\left(\sfrac13(\tilde{h}-\tilde{T})\right)\left(\frac{d\tilde{h}}{d\tilde{T}}-1\right)
+ \tilde{T} + \frac32\sin\left(\sfrac23(\tilde{h}-\tilde{T})\right)  = 0.
\end{equation}
Representing $\tilde{h}(\tilde{T})$ by a formal power series and inserting
this into~\eqref{gT} yields that $\theta = \frac13(-\tilde{T} +
\tilde{h}(\tilde{T}))$ can be written as
\begin{equation}\label{CM_Expansion2}
\theta = -\frac13\tilde{T}\left\{1 - \underbrace{\frac{1}{3^3}}_{0.037}\tilde{T}^2 +
\underbrace{\frac{2}{3^3\cdot 5}}_{0.015}\tilde{T}^4 - \underbrace{\frac{50}{3^6\cdot 7}}_{0.010}\tilde{T}^6
+ \underbrace{\frac{532}{3^{10}}}_{0.009}\tilde{T}^8\right\} + {\cal O}(\tilde{T}^{11}),
\end{equation}
when $\tilde{h}(\tilde{T})$ is expanded to 10th order.
Figure~\ref{fig:tildeT} shows that the center manifold expansion in
$\tilde{T}$ converges for small $T$ ($\tilde{T}<1\, \rightarrow\, T<1/2$),
but not for large $T$ (thus this formulation gives a center manifold
expansion with smaller range of convergence than the previous results in
$T$).

\begin{figure}[ht!]
\begin{center}
\subfigure[Pad\'e approximants in ${\bf S}$.]{\label{fig3:first}
\includegraphics[width=0.45\textwidth]{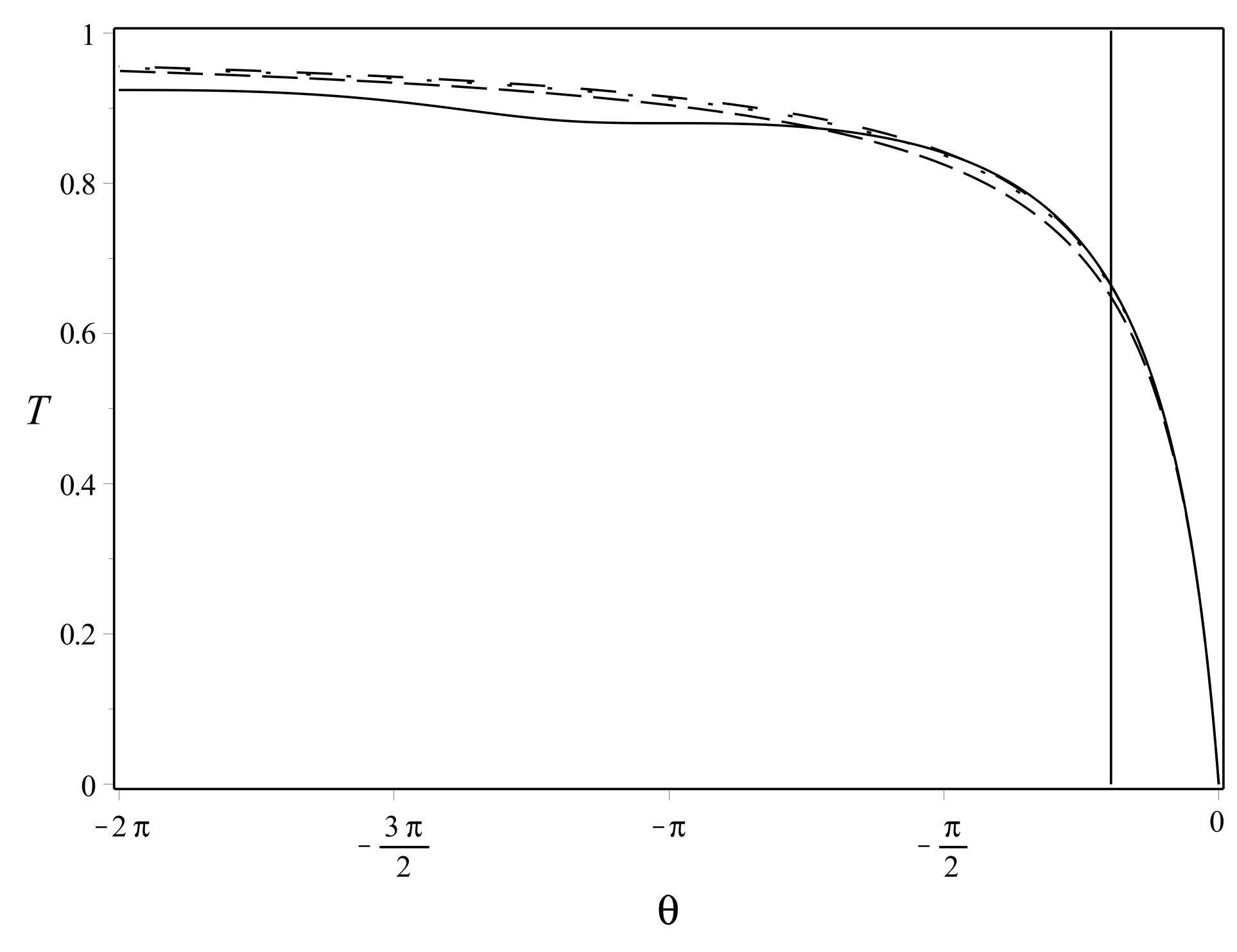}
}%
\subfigure[Pad\'e approximants near $q=0$.]{\label{fig3:second}
\includegraphics[width=0.45\textwidth]{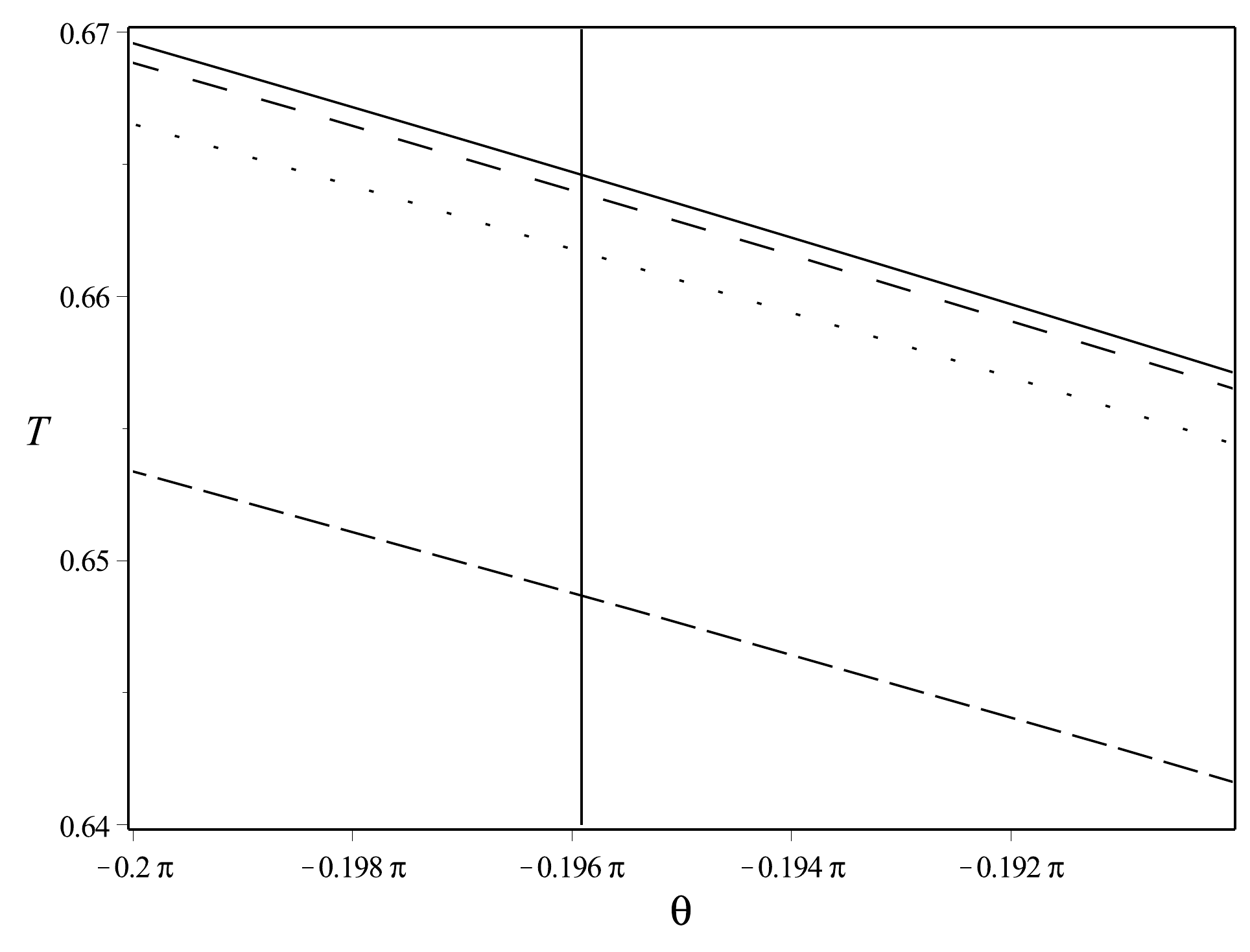}
}
\subfigure[Relative errors of Pad\'e approximants for small $\theta$]{\label{fig3:Padeerror}
\includegraphics[width=0.45\textwidth]{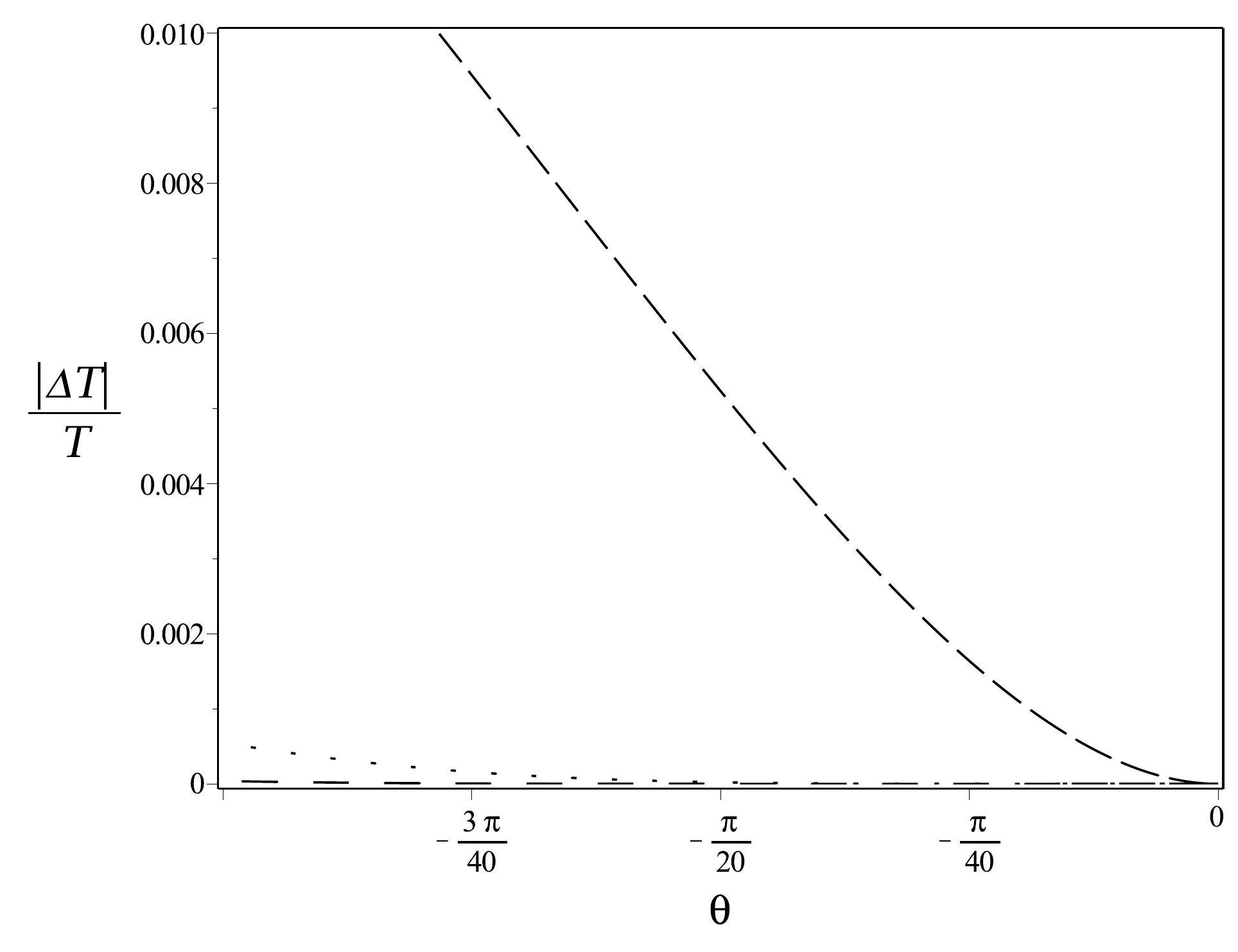}
}
\end{center}
\caption{The plot~\ref{fig3:first} shows the numerical solution (solid line) and the
$[1/1]_\theta$ (dashed), $[3/3]_\theta$ (space-dotted), and $[5/5]_\theta$ (space-dashed)
Pad{\'e} approximants in the state space ${\bf S}$. The plot~\ref{fig3:second}
depicts the situation close to $q=0$ (the vertical line) in more detail. The
plot~\ref{fig3:Padeerror} shows the relative errors for the Pad{\'e} approximants.}
\label{fig:Pade}
\end{figure}

Expressing~\eqref{CM_Expansion2} in $T$ yields a rational function, which
when expanded to 10th order precisely gives~\eqref{CM_Expansion}, which
converges for larger values than $T=1/2$. Next we turn to Pad{\'e}
approximants. Writing~\eqref{CM_Expansion2} as $\theta =
-\frac13\tilde{T}f(\tilde{T}^2)$ and writing $f(\tilde{T}^2)$ in terms of its
Pad{\'e} approximants for the various orders yield
\begin{subequations}\label{tildeTpade}
\begin{align}
[1/1]_{\theta} 
                &= -\frac{\tilde{T}}{3}f(0) = -\frac{\tilde{T}}{3}, \\
[3/3]_{\theta} 
                  & = -\frac{\tilde{T}}{3}[1/1]_{f} = -\frac{\tilde{T}}{3}\left(\frac{1+\frac{7^2}{3^3\cdot5}\tilde{T}^2}{1+\frac{2}{5}\tilde{T}^2}\right), \\
[5/5]_{\theta} 
                  &= -\frac{\tilde{T}}{3}[2/2]_{f} = -\frac{\tilde{T}}{3}\left(\frac{1+\frac{61\cdot419}{3^4\cdot13\cdot19}\tilde{T}^2+
                  \frac{2^4\cdot167\cdot1609}{3^4\cdot5\cdot7\cdot13\cdot19}\tilde{T}^4}{1+\frac{2^2\cdot5^2\cdot263}{3^4\cdot13\cdot19}\tilde{T}^2
                  +\frac{2\cdot5\cdot3659}{3^4\cdot7\cdot13\cdot19}\tilde{T}^4}\right),
\end{align}
\end{subequations}
where the notation on the right hand side corresponds to the Pad{\'e}
expression for $\theta(T)$ obtained from the series
expansion~\eqref{CM_Expansion2}. Thus using the center manifold expansion
associated with $\mathrm{dS}_+$ for the dynamical system~\eqref{dynsys2} is a
computationally convenient way of obtaining the `natural' convergent series
of Pad{\'e} approximants in a compressed form for the center manifold of
$\mathrm{dS}_+$ of the dynamical system~\eqref{dynsys}. In
eq.~\eqref{tildeTpade} each expression gives a curve in the state space ${\bf
S}$ that approximates the attractor solution, when $\tilde{T}$ is replaced by
$T$. The numerical solution and the above Pad{\'e} approximants are plotted
together with relative errors, $|\Delta T|/T = |T_\mathrm{Pad} -
T_\mathrm{num}|/T_\mathrm{num}$, in Figure~\ref{fig:Pade}. In addition we
give the relative errors $|\Delta T|/T$ and $|\Delta H|/H$ of the center
manifold Pad\'e approximants at the end of the accelerating regime, i.e., at
$q=0$, in Table~\ref{tab:padeTtheta}. As can be seen, increasingly higher
Pad{\'e} approximants give better results, indicating desirable convergence
and range properties. Moreover, note that the Pad\'e approximants yield much
better results than the associated series expansions described in
Section~\ref{sec:centexp}.
\begin{table}[ht!]
\begin{center}
 \begin{tabular}{|c|c|c|c|c|c|} \hline
                        &             &          &          &           \\ [-2ex]
                        & Num.        &   $[1/1]_\theta$  &  $[3/3]_\theta$   &  $[5/5]_\theta$    \\[1ex] \hline
                        &             &          &          &           \\ [-2ex]
   $T$                  & 0.6646      &  0.6487  &  0.6617  &  0.6639   \\[1ex] \hline
                        &             &          &          &           \\ [-2ex]
$\frac{|\Delta T|}{T}$  & \textemdash &  2.392\% &  0.436\% &  0.102\%  \\[1ex] \hline
                        &             &          &          &           \\ [-2ex]
$\frac{H}{m}$           &    0.5047   &  0.5416  &  0.5113  &  0.5062   \\[1ex] \hline
                        &             &          &          &            \\ [-2ex]
$\frac{|\Delta H|}{H}$  & \textemdash &  7.308\% &  1.299\% &  0.298\%   \\ [1ex] \hline
 \end{tabular}
\end{center}
\caption{Numerical values and relative errors for the Pad\'e approximants at
$q=0$.}\label{tab:padeTtheta}
\end{table}
%

\subsection{Global approximations}

\begin{figure}[ht!]
\begin{center}
{\subfigure[Global piecewise Pad{\'e}-averaged approximation.]{
        \label{fig:avematched}
        \includegraphics[width=0.45\textwidth]{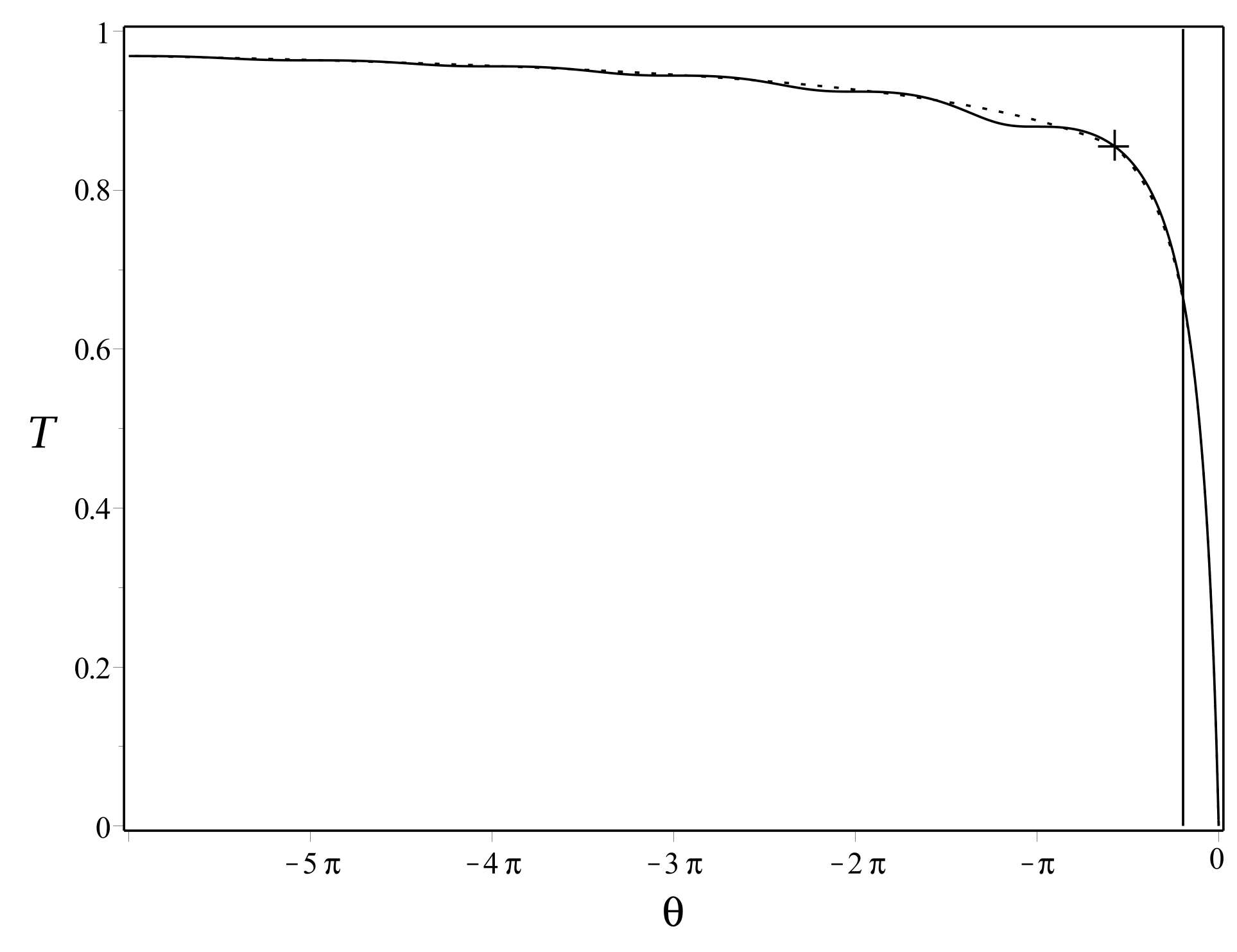}}}
{\subfigure[Matched global Pad{\'e}-oscillatory approximation.]{
        \label{fig:oscmatched}
        \includegraphics[width=0.45\textwidth]{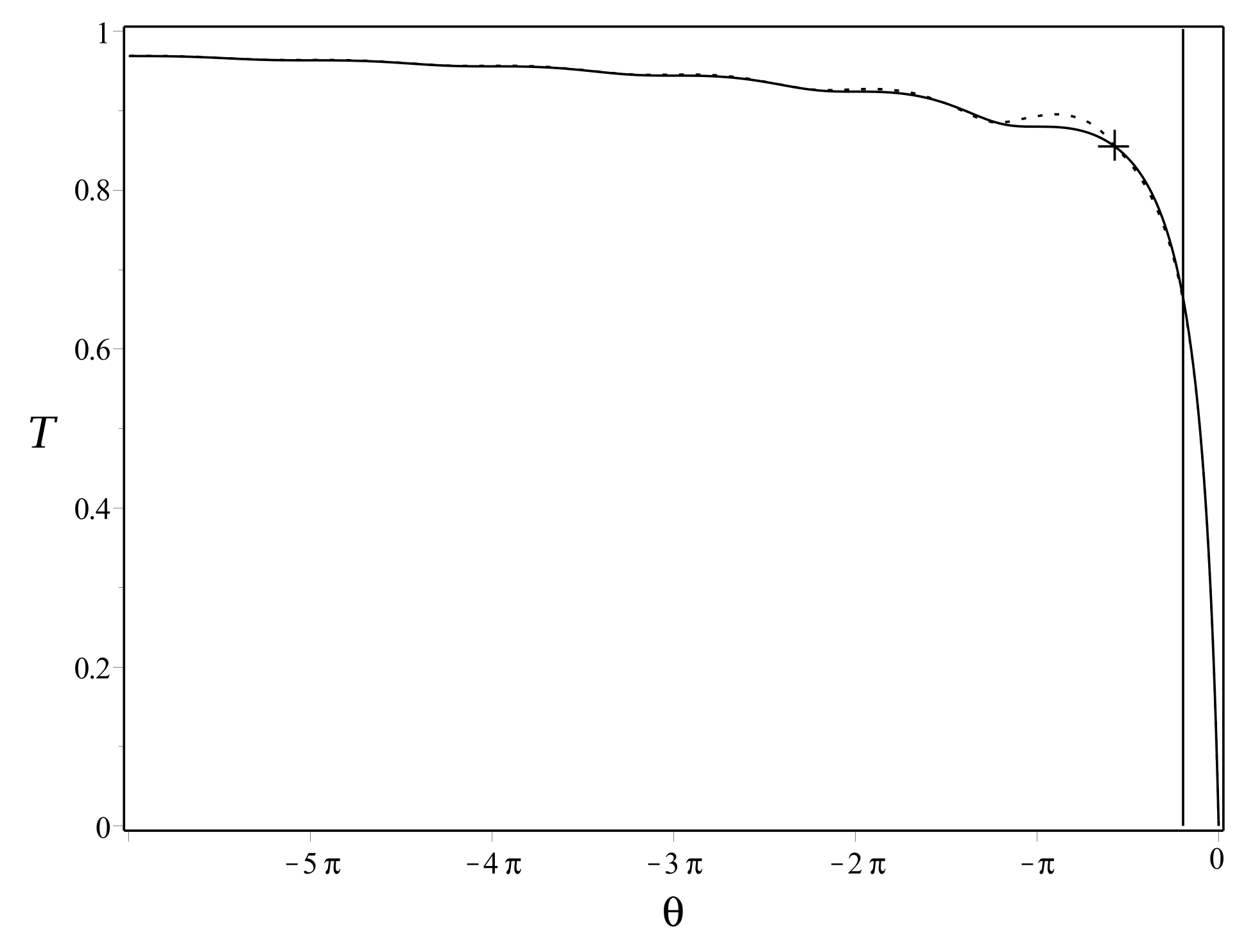}}}
\end{center}
        \caption{The plot~\ref{fig:avematched} shows the numerical
solution (solid line) and the $[3/3]_\theta$ Pad{\'e} approximant matched (at the cross) with
the averaged solution where the $[3/3]_\theta$ Pad{\'e} approximant crosses
the numerical solution at $T\approx 0.8552$. The plot~\ref{fig:oscmatched} shows the numerical
solution (solid line) and the $[3/3]_\theta$ Pad{\'e} approximant matched
with the approximate oscillatory solution for late times (at the cross, where
$T \approx 0.8552,\,t_0 \approx -2.1212$).}
    \label{fig:averagematched}
\end{figure}

Thanks to the large range of the center manifold based Pad{\'e} approximants
these can be joined with the approximations for the evolution at late times
to yield temporally global approximations for the attractor solution. In
Figure~\ref{fig:avematched} the $[3/3]_{\theta}$ approximant, given in
eq.~\eqref{tildeTpade}, is matched with the average approximation for late
times, given in eq.~\eqref{thetaTlate}, to yield a global piecewise
approximation for the attractor solution. Figure~\ref{fig:oscmatched} depicts
a temporally global description of the attractor solution obtained by joining
the $[3/3]_\theta$ Pad{\'e} approximant with the late time approximation
given in eq.~\eqref{thetaTlateosc}.

Note that an approximate solution for the entire attractor solution, which
subsequently can be used to obtain an approximate solution in e.g.
$(\phi,\dot{\phi})$-space, may make it possible to at least approximately
obtaining or estimating a global measure~\cite{remcar13} in
$(\phi,\dot{\phi})$-space (or $(T,\theta)$-space for $T\in(0,1)$) that might
shed further light on the meaning of an `attractor solution.'

Finally, it is worth pointing out that one can also obtain approximations for
the remaining solutions that originate from the hyperbolic fixed points
$\mathrm{M}_\pm$ by series expansions of these fixed points based on Picard's
method, as described in the Bianchi type II perfect fluid case
in~\cite{ugg89}, see also references therein and~\cite{limetal04}. These
expansions can be used to obtain approximants in order to improve the rate
and range of convergence, and it is thereby possible to approximately
describe the entire solution space. A similar statement holds for a wide
range of related problems in General Relativity and modified gravity
theories; again, we stress that the current minimally coupled scalar field
with a quadratic potential just serves as an illustrative example.

\section{Slow-roll and center manifold approximant comparisons}\label{sec:slowroll}

The purpose with this section is twofold: (i) to use the present example with
a quadratic potential to clarify the slow-roll approximation and its
expansion extensions, showing that the so-called slow-roll approximation
actually corresponds to several approximations for the attractor solution
(this is a point that is related to the ones raised in the recent
paper~\cite{ven14} as regards so-called horizon-flow approximations), (ii) to
quantitatively assess the accuracy of slow-roll approximations for the
present quadratic scalar field potential. In forthcoming papers we will
generalize and contextualize these results to increasingly general
situations; in particular we will in a subsequent paper deal with monomial
potentials and perfect fluids as the source.

\begin{figure}
\begin{center}
{\subfigure[Hubble slow-roll approximants]{\label{fig:LiddleH}
\includegraphics[width=0.45\textwidth]{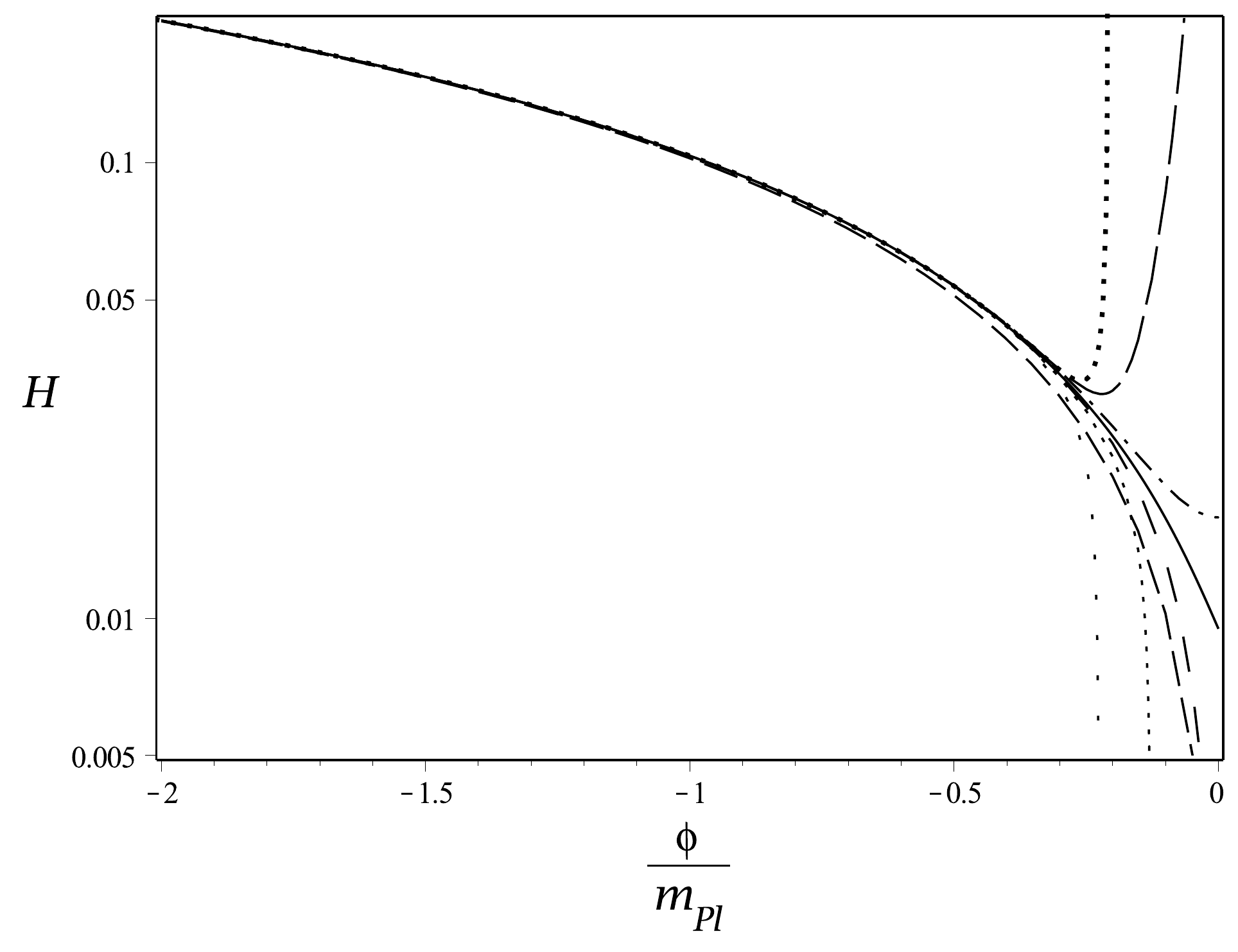}}}
{\subfigure[Hubble slow-roll errors]{\label{fig:Liddleerror}
\includegraphics[width=0.45\textwidth]{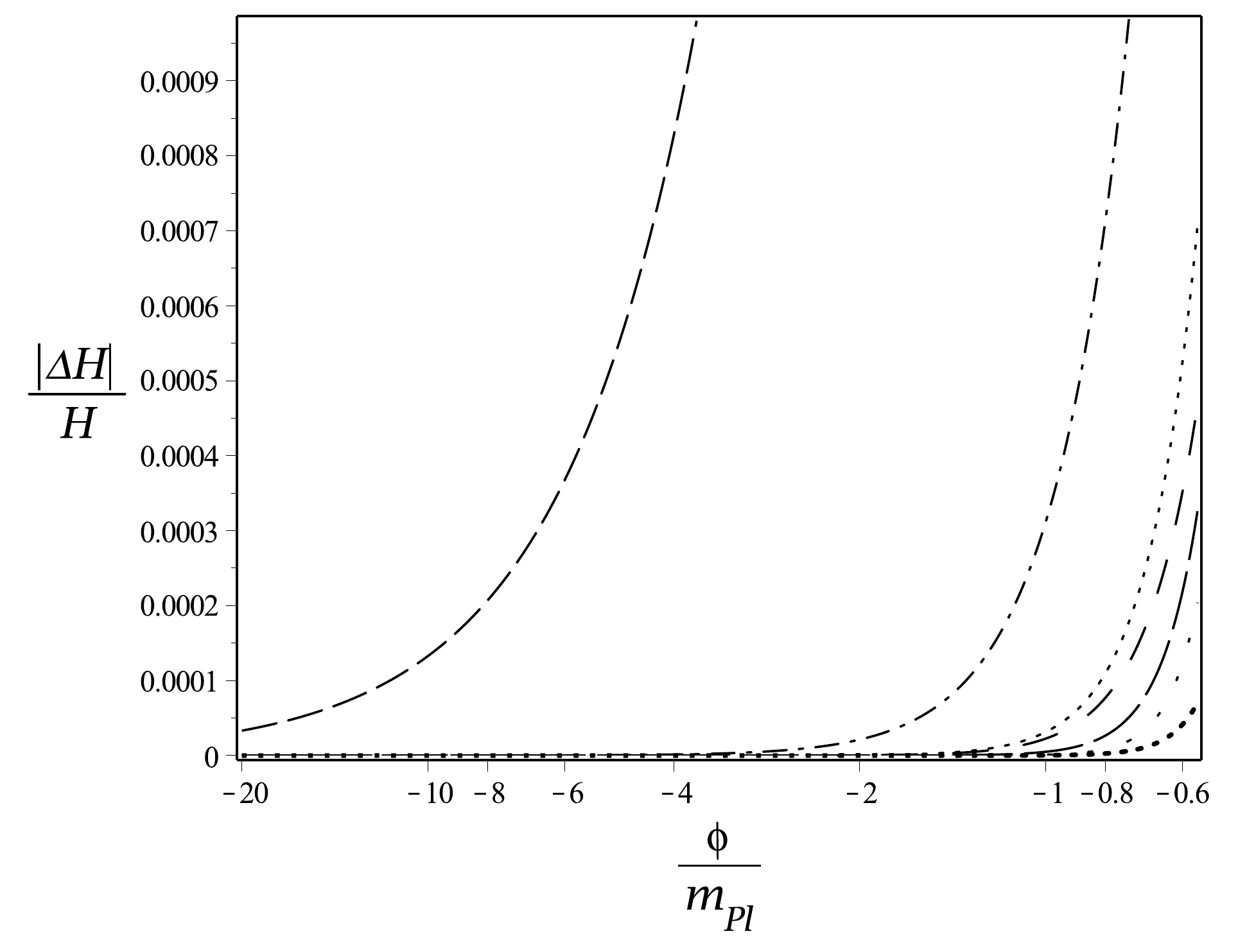}}}
\end{center}
\caption{The plot~\ref{fig:LiddleH} describes the numerical solution (solid line) and the Hubble
slow-roll approximants (in this plot $m$ has been chosen so that the
lower order approximants yield the results in~\cite{lidetal94}). The plot~\ref{fig:Liddleerror} gives the
relative errors of the Hubble slow-roll approximants for large values of $\phi$.
Both plots depict the 0th (dashed), 1st (dot-dashed), 2nd (dotted),
3rd (long-dashed), and 4th (space-dotted) order approximations, and
the $[1/1]$ (space-dashed) and $[2/2]$
(fat-dotted) Canterbury approximants.}
\label{fig:LiddleHError}
\end{figure}

\begin{figure}
\begin{center}
{\includegraphics[width=0.5\textwidth]{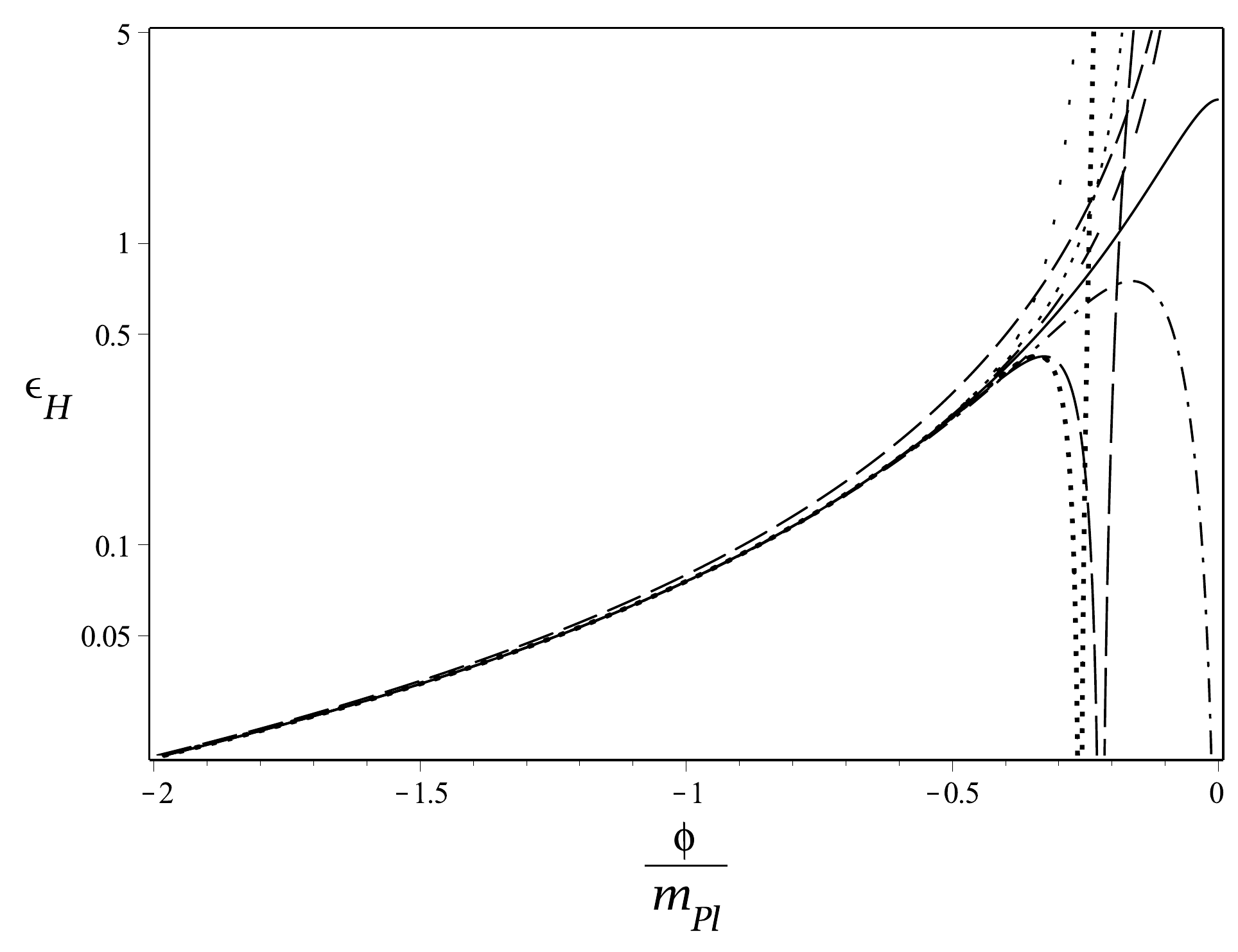}}%
\end{center}
\caption{$\epsilon_H(\phi/m_\mathrm{Pl})$: The notation for the various curves is the same as in the previous figures.
     }%
\label{fig:epsHphi}
\end{figure}

To discuss slow-roll approximants we first reproduce and extend the results
by Liddle {\it et al}.~\cite{lidetal94} who introduced a hierarchy of Hubble
slow-roll parameters, which were subsequently related to a hierarchy of
potential slow-roll parameters in order to produce analytic slow-roll
expansions and approximants. In terms of our state space variables the first
two Hubble slow-roll parameters are given by
\begin{equation}\label{slowrollparameters}
\epsilon_H = 1+ q = 3\sin^2\theta,\qquad \eta_H = 3 + \left(\frac{T}{1-T}\right)\cot\theta .
\end{equation}
To facilitate comparisons with~\cite{lidetal94} we here keep the coupling
constant $\kappa=8\pi m^{-2}_\mathrm{Pl}$. Following the methods described
in~\cite{lidetal94} results in that to 4th order for the present quadratic
potential the slow-roll expansion for $(H/m)^2$ in $\phi^2$ is given by
\begin{equation}\label{Hslowroll}
\left(\frac{H}{m}\right)^2=\frac{\kappa\phi^2}{6}
\left[1+\frac{2}{3\kappa\phi^{2}}-\frac{2^2}{3^2\kappa^2\phi^{4}} +
\frac{2^5}{3^3\kappa^3\phi^{6}} -\frac{2^4\cdot5^2}{3^4\kappa^4\phi^{8}}+{\cal O}\left(\frac{1}{\kappa^5\phi^{10}}\right)\right].
\end{equation}
while the first so-called Canterbury approximants, based on slow-roll
potential parameters as described in~\cite{lidetal94}, are given by
\begin{subequations}\label{HPade}
\begin{align}
[1/1] &=
\frac{\kappa\phi^2}{6}\left[\frac{1+\frac{2^2}{3\kappa\phi^{2}}}{1+\frac{2}{3\kappa\phi^{2}}}\right],\\
[2/2] &=
\frac{\kappa\phi^2}{6}\left[\frac{1+\frac{13}{2\kappa\phi^2} +
\frac{2\cdot5^2}{3^2\kappa^2\phi^4}}{1+\frac{5\cdot7}{2\cdot3\kappa\phi^2} +
\frac{19}{3^2\kappa^2\phi^4}-\frac{7\cdot2^{2}}{3^3\kappa^4\phi^8} -
\frac{2\cdot7\cdot13}{3^3\kappa^5\phi^{10}}-\frac{2^{3}\cdot5^2\cdot7}{3^5\kappa^6\phi^{12}}}\right].
\end{align}
\end{subequations}
The slow-roll expansions and Canterbury approximants are compared with the
numerical attractor solution in Figure~\ref{fig:LiddleHError}. For large
$\phi$, and thereby large $H$, there is convergence for the series
expansions, but this is no longer the case for small $\phi$. Furthermore,
note that the best approximant for small $\phi$ is the $[1/1]$ approximant.
The same holds for the plot of $\epsilon_H = -\frac{d\ln H}{d\ln a} =
3\dot{\phi}^2/(\dot{\phi}^2 + m^2\phi^2)$ versus $\phi/m_\mathrm{Pl}$ given
in Figure~\ref{fig:epsHphi}.

In the present context, it follows from eq.~\eqref{slowrollparameters} that
the slow-roll limit $\epsilon_H=\eta_H=0$ corresponds to the fixed points
$\mathrm{dS}_\pm$ and that only imposing `the attractor condition' $\eta_H=0$
close to (without loss of generality) $(T,\theta)=(0,0)$ yields
$T+3\theta=0$, i.e., the attractor condition leads to the tangent space $E^c$
of the center manifold solution in the slow-roll limit. It therefore follows
that the slow-roll expansions in~\cite{lidetal94} corresponds to an analytic
attempt to describe the center manifold, i.e., the attractor solution, to
increasing accuracy, which in turn describes the intermediate time behavior
of an open set of solutions that passes close to the de Sitter fixed point
(where measures attempt to describe how `large' this open set is).

\begin{figure}[ht!]
\begin{center}
{\subfigure[Hubble slow-roll approximations.]{\label{fig:HSR}
\includegraphics[width=0.45\textwidth]{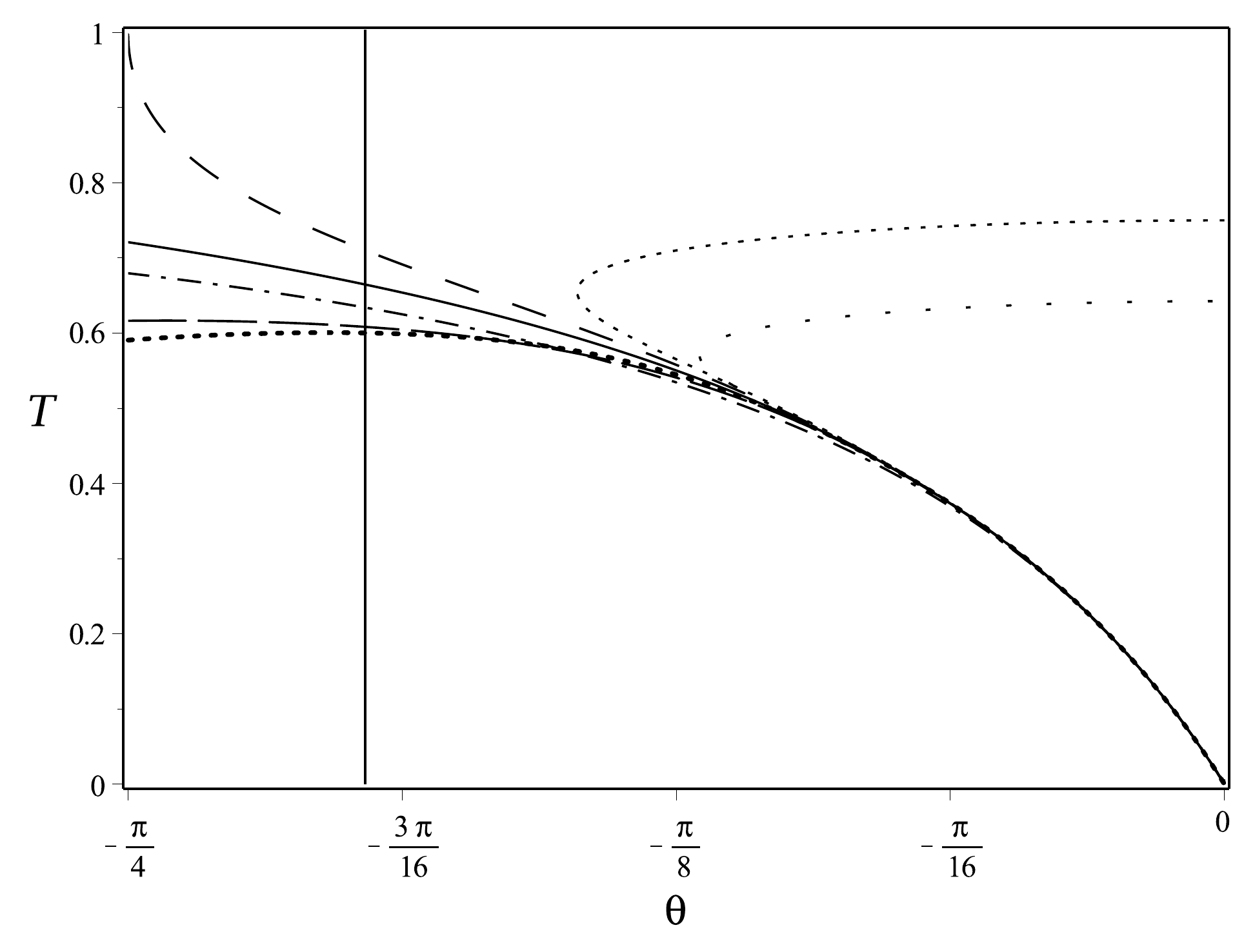}}}
{\subfigure[Relative errors for Hubble slow-roll approximations.]{\label{fig:HSRerrors}
\includegraphics[width=0.45\textwidth]{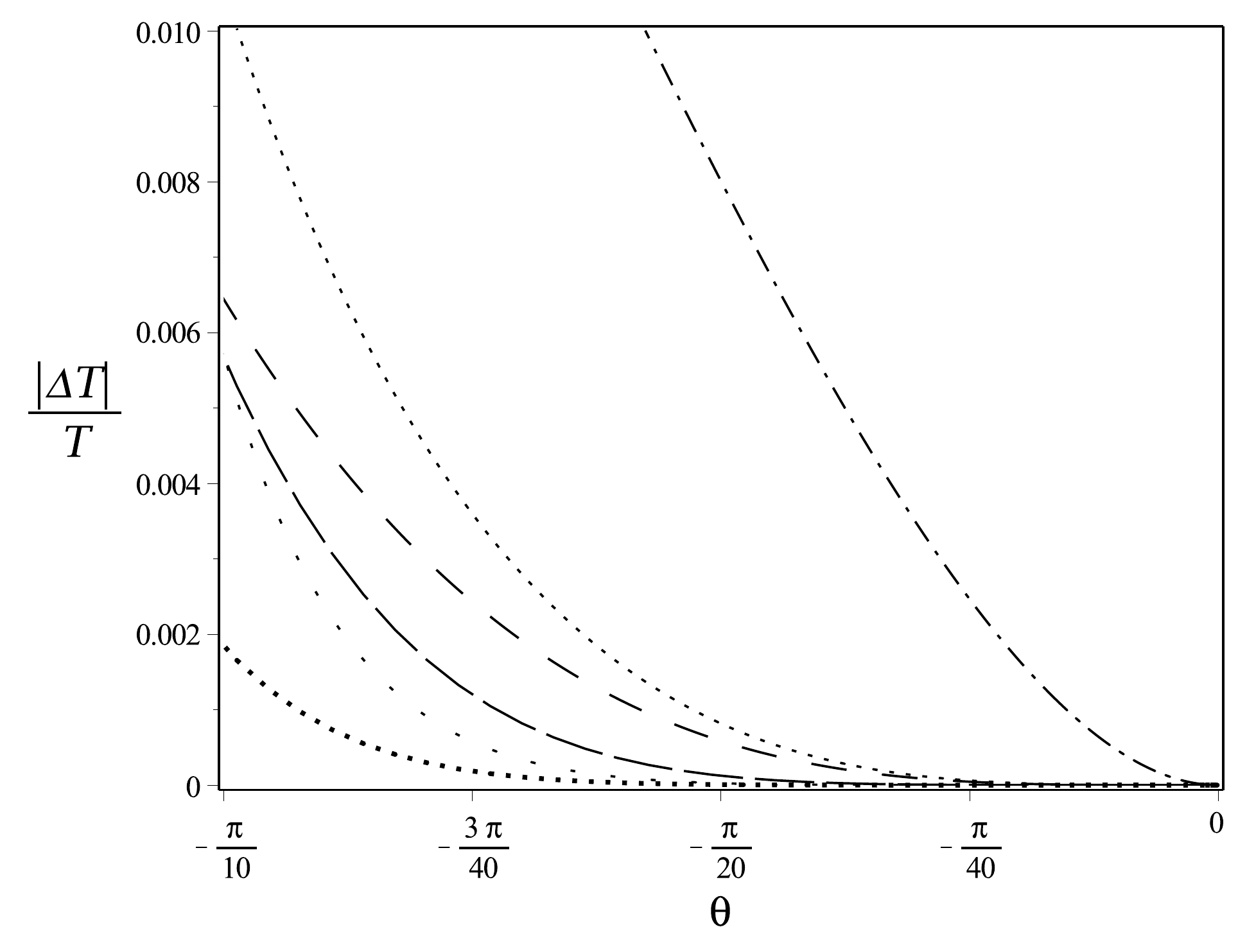}}}
\end{center}
\caption{The plot~\ref{fig:HSR} describes the numerical solution (solid line) and
the different Hubble slow-roll approximations: the 1st (dot-dashed), 2nd (dotted),
3rd (long-dashed), and 4th (space-dotted) order approximations, and the [1/1] (space-dashed)
and [2/2] (fat-dotted) Canterbury approximants. The plot~\ref{fig:HSRerrors} gives
the relative errors for the Hubble slow-roll approximations.}%
\label{fig:Hslowroll}
\end{figure}

Slow-roll approximations actually consist of two classes of approximations.
The first class of approximations arises from obtaining approximate
expressions for $H^2(\phi)$. These are the so-called Hubble slow-roll
expansions given in eq.~\eqref{Hslowroll} and the associated Canterbury
approximants given in eq.~\eqref{HPade}, which give rise to a family of
curves in ${\bf S}$ which yield approximations for the attractor solution.
These approximation curves are obtained by performing the
transformation~\eqref{Newvar} (after having set $8\pi m_\mathrm{Pl}^{-2}=1$).
The curves and relative errors for various series expansions and approximants
are given in Figure~\ref{fig:Hslowroll} together with the numerically
calculated attractor solution; in addition we describe numerical values and
relative errors at $q=0$ in Table~\ref{tab:slowrollq0}. Note that to zeroth
order the Hubble expansion gives $\cos^2\theta=1$, which corresponds to the
straight lines in $\bar{\bf S}$ describing $q=-1$.
\begin{table}[ht!]
\begin{center}
\begin{tabular}{|c|c|c|c|c|c|c|c|c|} \hline
                              &             &             &          &             &          &          &              &           \\ [-2ex]
                              &    Num.     &   0th       &   1st    &     2nd     &   [1/1]  &   3rd    &      4th     &   [2/2]   \\ [1ex]\hline
                              &             &             &          &             &          &          &              &           \\ [-2ex]
         $T$                  &   0.6646    & \textemdash & 0.6340   & \textemdash & 0.7101   & 0.6083   &  \textemdash & 0.6002    \\  [1ex]\hline
                              &             &             &          &             &          &          &              &           \\ [-2ex]
$\frac{|\Delta T|}{T}$  & \textemdash & \textemdash & 4.604\%  & \textemdash & 6.846\%  & 8.471\%  &  \textemdash & 9.690\%   \\ [1ex]\hline
                              &             &             &          &             &          &          &              &           \\ [-2ex]
$\frac{H}{m}$                 & 0.5047      & \textemdash & 0.5774   & \textemdash & 0.4082   & 0.6440   &  \textemdash & 0.6660    \\ [1ex]\hline
                              &             &             &          &             &          &          &              &           \\ [-2ex]
$\frac{|\Delta H|}{H}$        & \textemdash & \textemdash & 14.400\% & \textemdash & 19.111\% & 27.598\% &  \textemdash & 31.952\%  \\ [1ex]\hline
 \end{tabular}
\end{center}\label{tab:Hubbleslowroll}
\caption{Numerical values and relative errors for Hubble slow-roll expansions
and Canterbury approximants at $q=0$.}\label{tab:slowrollq0}
\end{table}

As can be seen from Figure~\ref{fig:Hslowroll}, the Hubble series expansion
converge for small $T$ (i.e., large $H$), which is the slow-roll regime, but
not for large $T$ (i.e. small $H$), as can be expected. The [2/2] Canterbury
approximant is the best approximation for small $T$, while the 1st order
Hubble expansion approximation is the most accurate among the slow-roll
approximants for large $T$ at $q=0$, see Figure~\ref{fig:Hslowroll} and
Table~\ref{tab:slowrollq0}. This is to be contrasted with the center manifold
expansions and the odd Pad{\'e} approximants which converge even beyond the
inflationary stage, far from where the slow-roll conditions break down. It is
this larger range of convergence that makes it possible to obtain quite good
piecewise global approximations for the attractor solution by matching center
manifold based approximants with approximations for the evolution at late
times, as shown in the previous section.

\begin{figure}
\begin{center}
{\subfigure[Relative errors: 1st order Hubble expansion approximation and center manifold approximants]{\label{fig:slowrollcomp}
\includegraphics[width=0.45\textwidth]{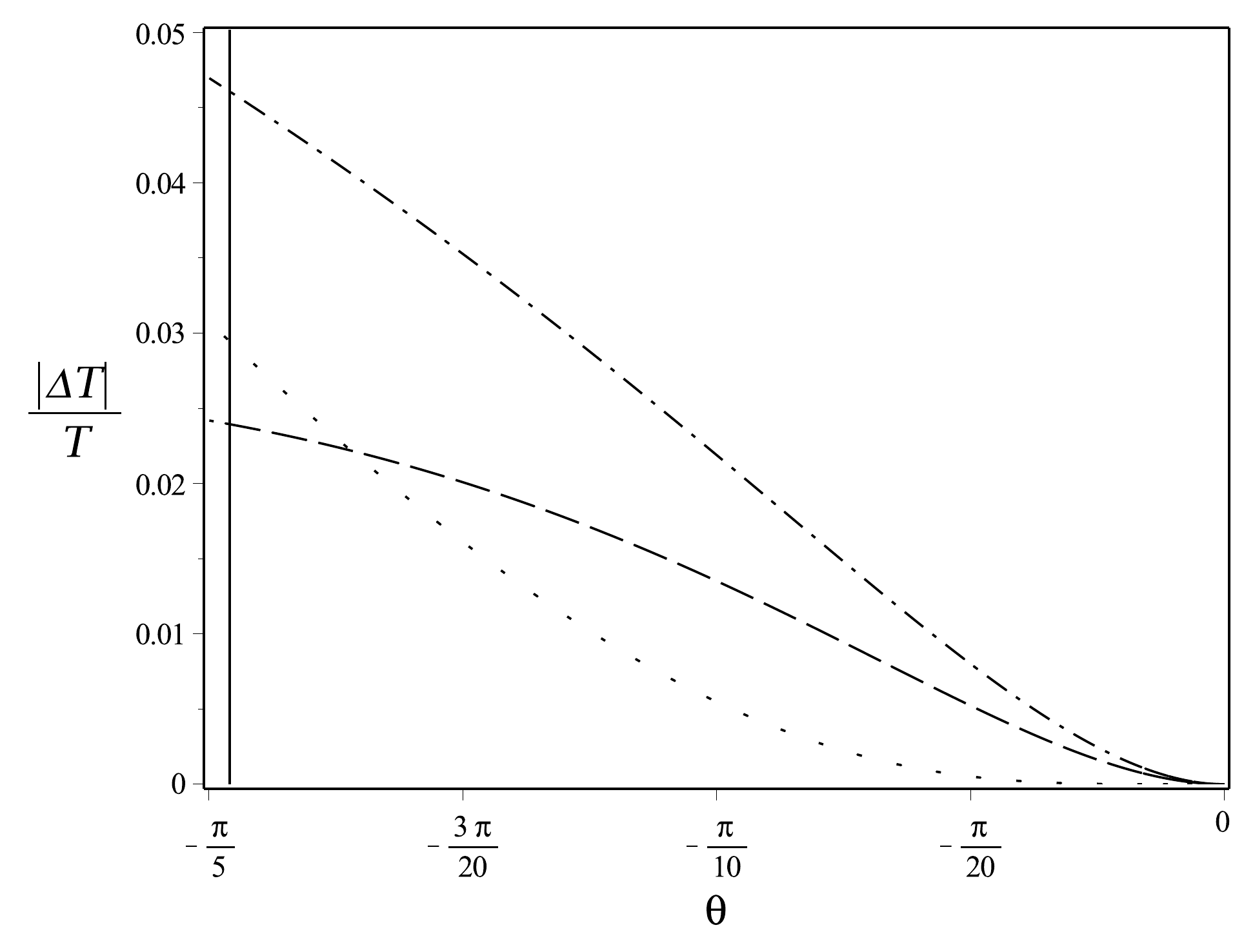}}}
{\subfigure[Relative errors: Hubble expansion Canterbury approximant and center manifold approximants]{\label{fig:cantcomp}
\includegraphics[width=0.45\textwidth]{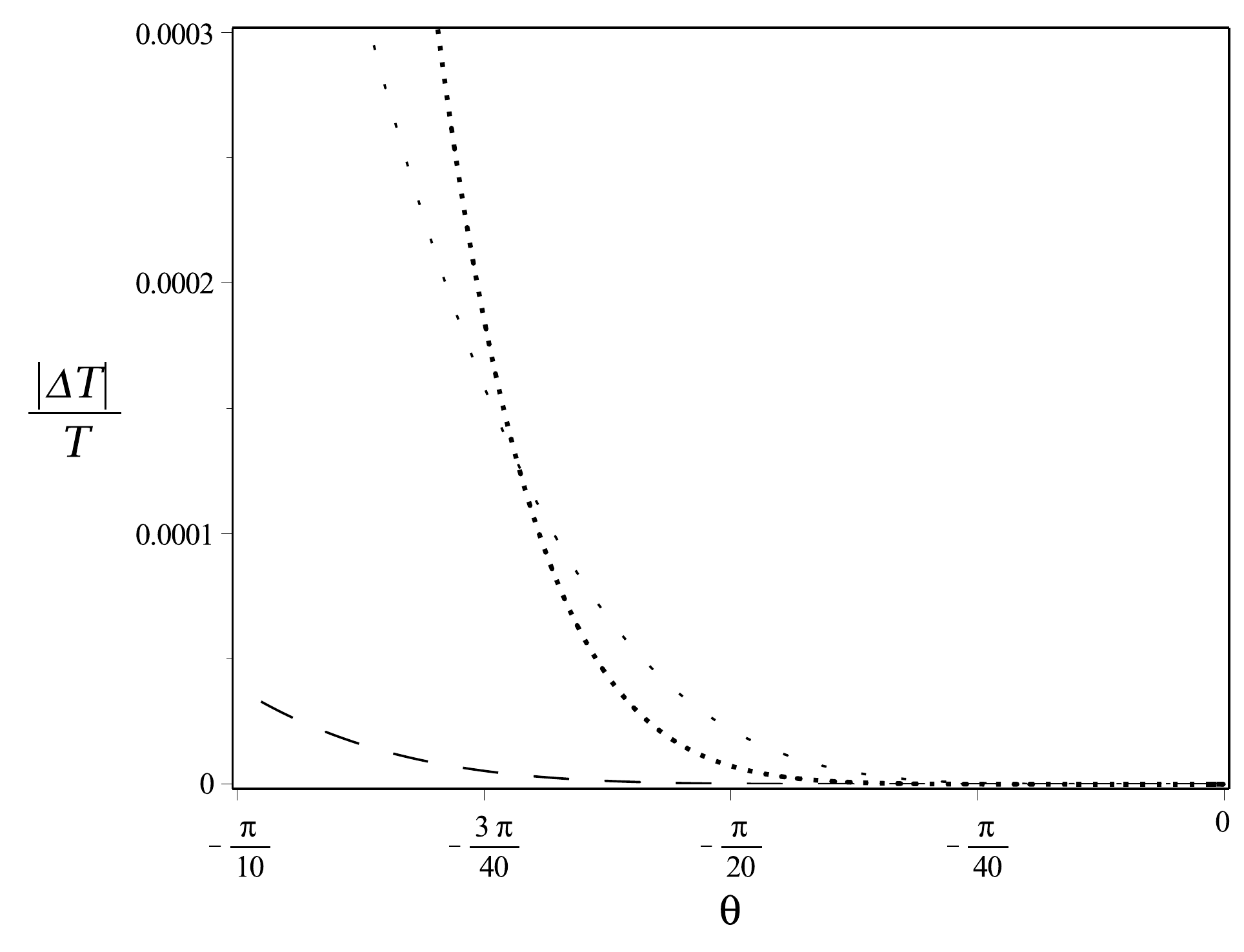}}}
\end{center}
\caption{The plot~\ref{fig:slowrollcomp} describes the relative errors for the the 1st
order Hubble expansion approximation (dash-dotted);
the 6th order center manifold expansion (space-dotted); the $[1/1]_\theta$ Pad{\'e} approximant (dashed).
The plot~\ref{fig:cantcomp} describes the relative errors of the $[2/2]$ Canterbury approximant (fat dotted);
the $[3/3]_\theta$ Pad{\'e} approximant (space-dotted); the $[5/5]_\theta$ Pad{\'e} approximant (space-dashed).}
\label{fig:slowrollbeaten}
\end{figure}

When comparing the 1st order Hubble expansion approximant with the center
manifold approximants in Figure~\ref{fig:slowrollcomp} it is seen that the
6th order center manifold expansion and the $[1/1]_\theta$ Pad{\'e}
approximant yields a more accurate approximation for the attractor solution
than the 1st order Hubble expansion approximation everywhere, although
especially for large $|\theta|$ and $T$. Furthermore, in
Figure~\ref{fig:cantcomp} it is shown that the $[5/5]_\theta$ Pad{\'e}
approximant gives a better approximation for the attractor solution than the
$[2/2]$ Canterbury approximant everywhere, especially for large $|\theta|$
and $T$, while the $[3/3]_\theta$ Pad{\'e} approximant is better than the
$[2/2]$ Canterbury approximant everywhere except for quite small $|\theta|$
and $T$.

The second class of slow-roll approximations for the attractor solution
arises from inserting the first class of Hubble expansion and Canterbury
approximants into
\begin{equation}\label{phidotHphi}
\dot{\phi} = - 2\frac{\partial H}{\partial \phi}
\end{equation}
(obtained by using $\phi$ locally as the independent variable and using the
present units). The usual slow-roll aproximation is obtained by inserting the
zeroth order Hubble expansion approximation into the above
equation~\cite{salbon90}, which for the present models yields the same
expression as the 1st order Hubble expansion approximation given previously
(cf. Tables~\ref{tab:slowrollq0} and~\ref{tab:phidotrelerr}) and which we
have compared with center manifold approximants in
Figure~\ref{fig:slowrollcomp}. (For the present models, the 1st $\dot{\phi}$
slow-roll approximation also gives the $[1/1]$ Hubble slow-roll
approximation.)

\begin{figure}
\begin{center}
\includegraphics[width=0.45\textwidth]{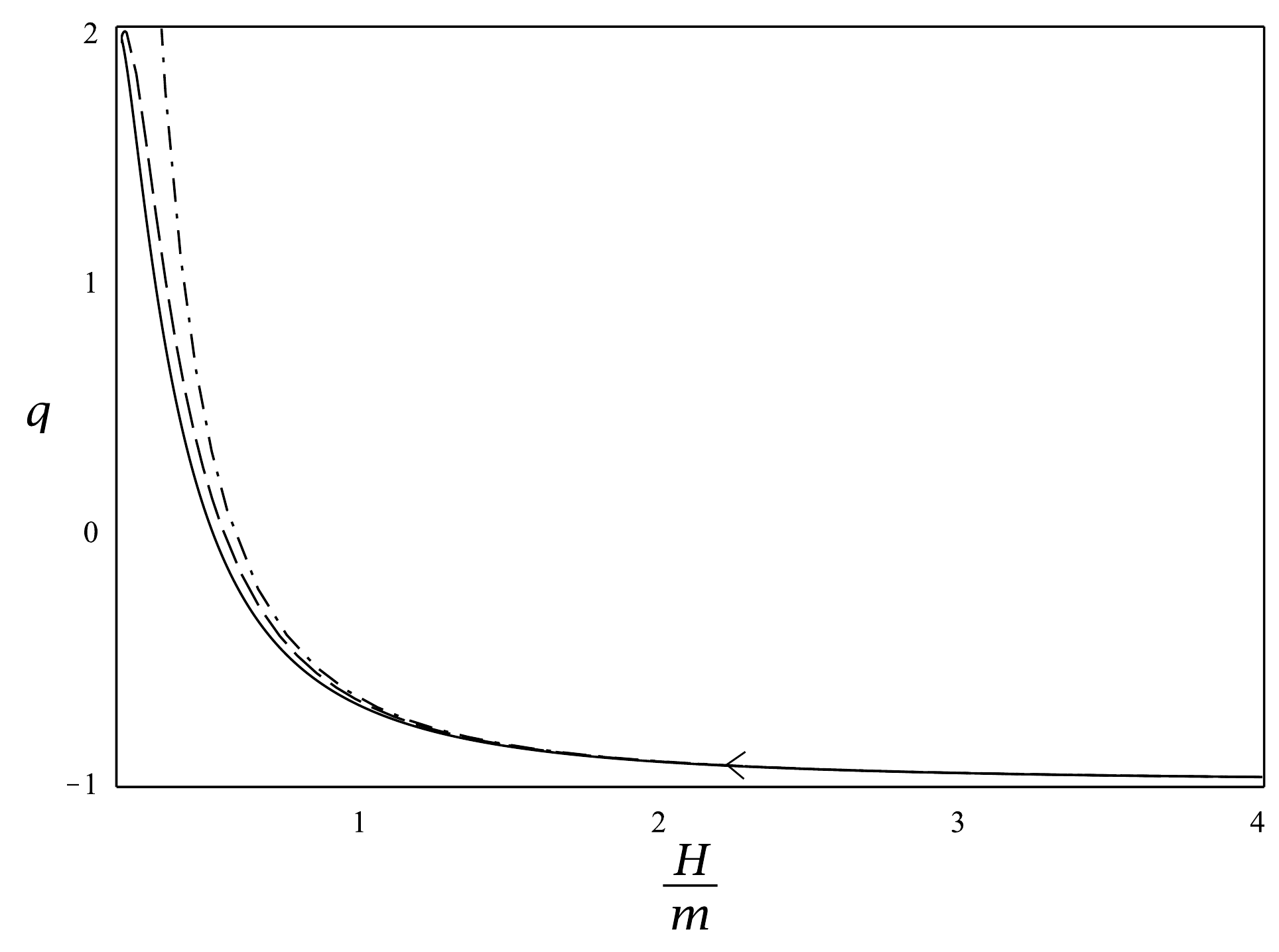}
\end{center}
\caption{The plot describes $q(H/m)$ for the slow-roll approximation
(dash-dotted); the $[1/1]_\theta$ Pad{\'e} approximant (dashed); and the numerical attractor solution (solid line).}
\label{fig:Approximationsq}
\end{figure}

It is of interest to note that the slow-roll approximation and the
$[1/1]_\theta$ Pad{\'e} approximant yield the following expressions for the
deceleration parameter $q$ as a function of $H/m$:
\begin{subequations}
\begin{alignat}{2}
q &= -1 + \frac13\left(\frac{m}{H}\right)^2 &\qquad & (\text{slow-roll}),\\
q &= - 1 + 3\sin^2\left[\frac13\left(\frac{m}{H}\right)\right] &\qquad &
([1/1]_\theta\,\, \text{Pad\'e}).
\end{alignat}
\end{subequations}
These expressions have been plotted in Figure~\ref{fig:Approximationsq}
together with the attractor solution.

\begin{figure}[ht!]
\begin{center}
\subfigure[$\dot{\phi}$ slow-roll approximations.]{\label{fig:PhiDotSR}
\includegraphics[width=0.45\textwidth]{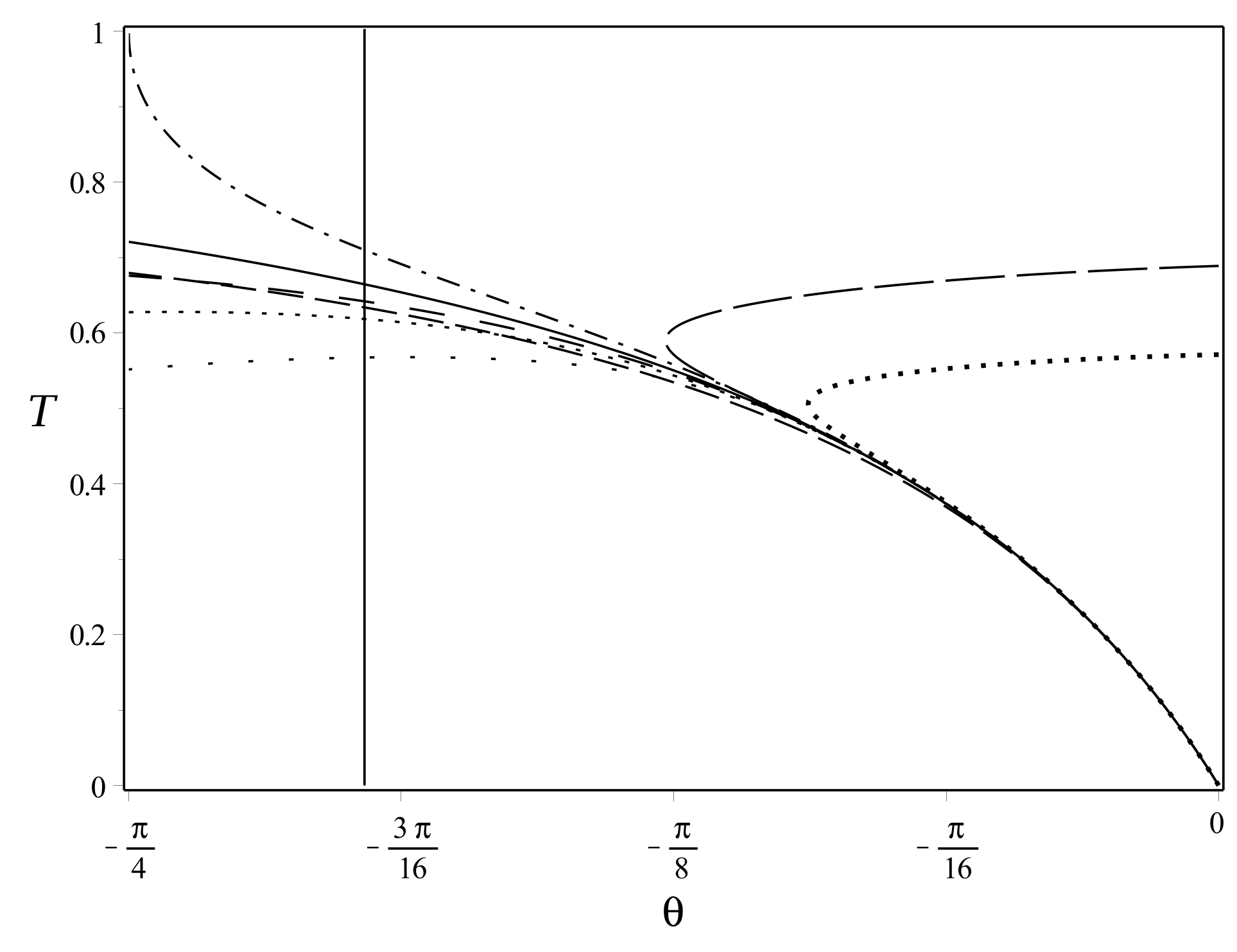}
}%
\subfigure[Relative errors for the $\dot{\phi}$ slow-roll approximations.]{\label{fig:PhiDotSRerrors}
\includegraphics[width=0.45\textwidth]{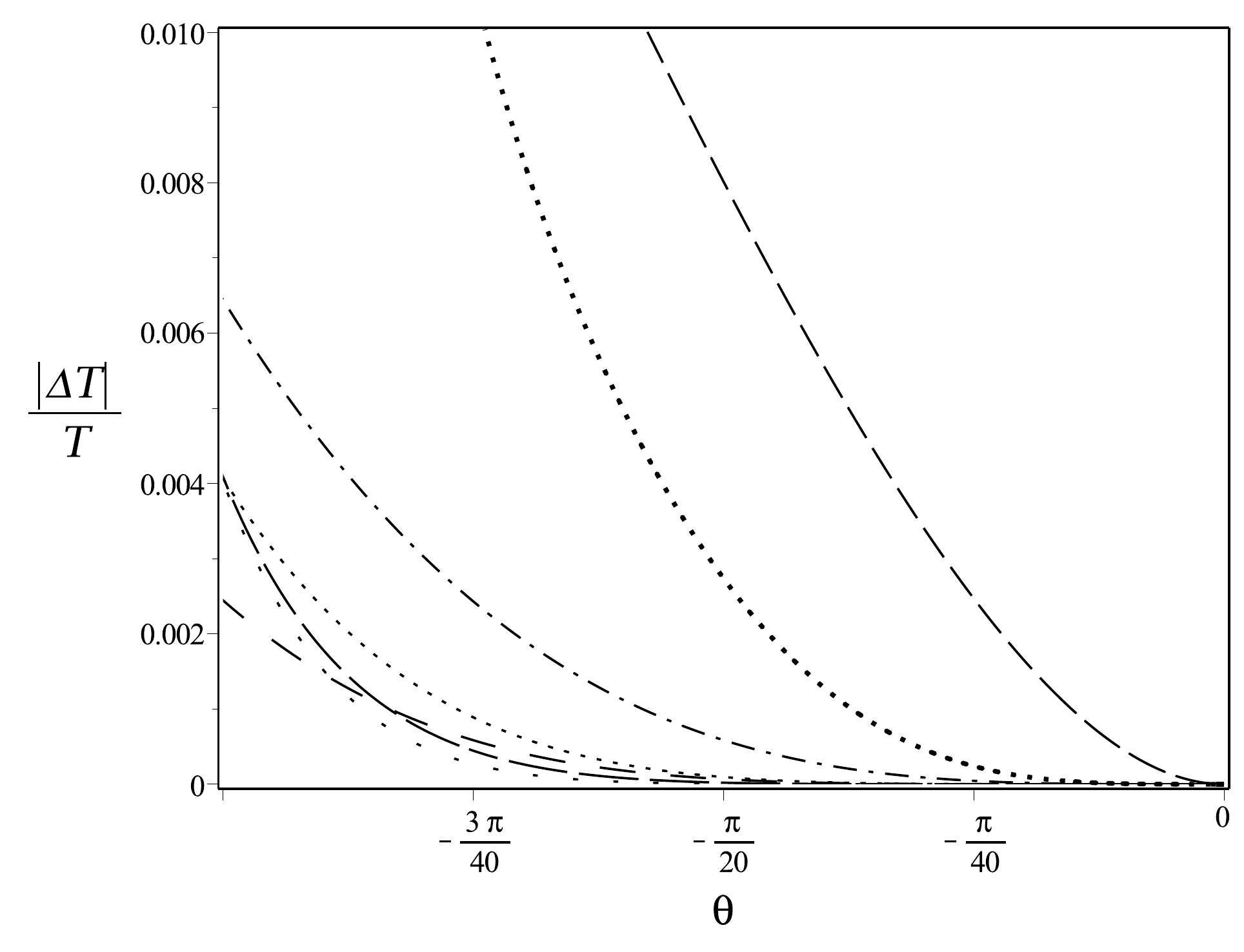}
}
\end{center}
\caption{Numerical solution (solid line) and $\dot{\phi}$ slow-roll approximations
to 0th (dashed), 1st (dot-dashed), 2nd (dotted), 3rd (long-dashed), and 4th (space-dotted)
order approximation, and the [1/1] (space-dashed) and [2/2] (fat-dotted) Canterbury approximants.}
\label{fig:phidotexp}
\end{figure}

\begin{figure}
\begin{center}
\includegraphics[width=0.45\textwidth]{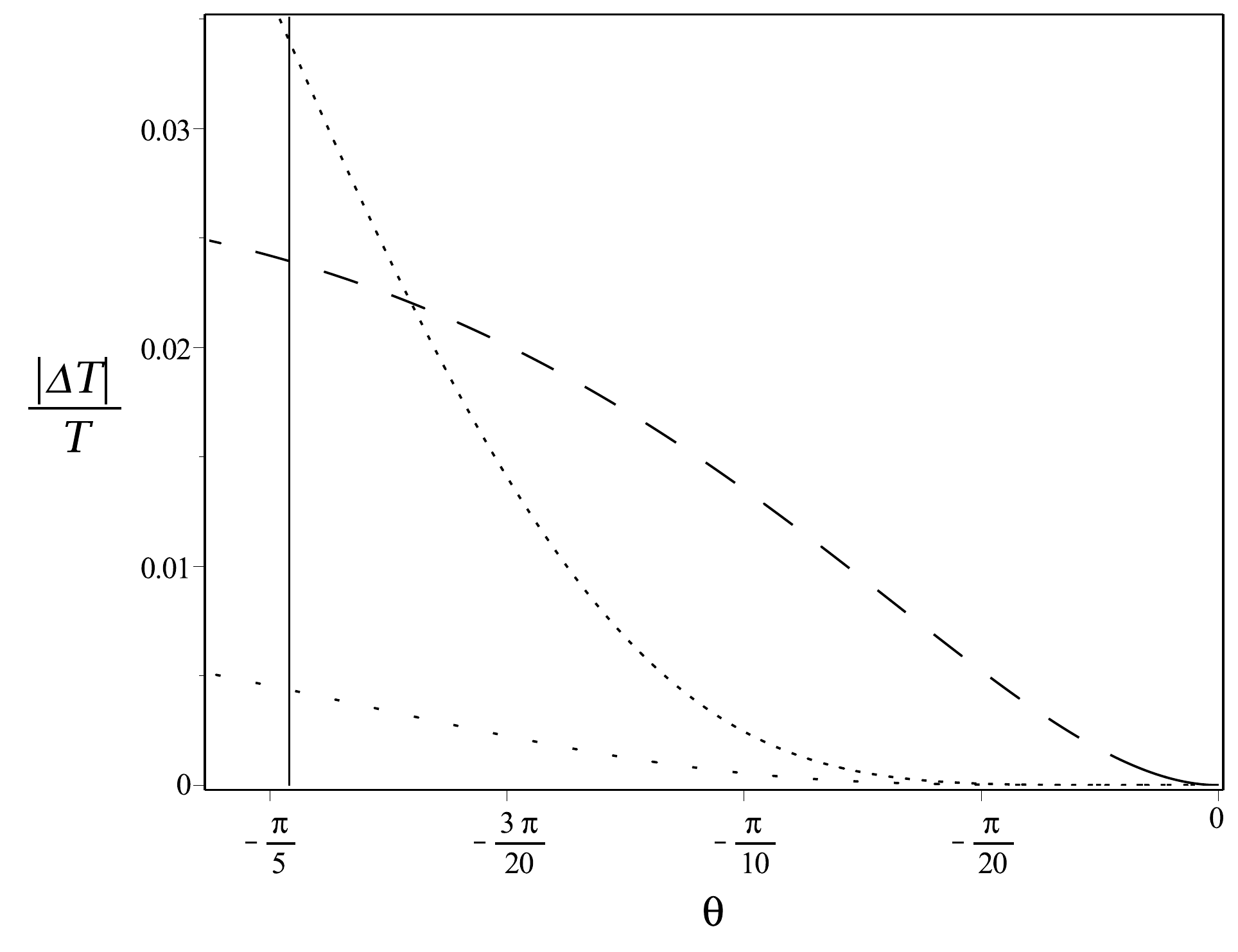}
\end{center}
\caption{The plot describes the relative errors for the $[1/1]$ Canterbury $\dot{\phi}$ slow-roll approximation
(dotted); the $[1/1]_\theta$ Pad{\'e} approximant (space-dashed); the $[3/3]_\theta$ Pad{\'e} approximant (space-dotted).}
\label{fig:2phidotcomp}
\end{figure}

As shown in Figure~\ref{fig:phidotexp}, the approximations based
on~\eqref{phidotHphi} are more accurate than the Hubble expansions for small
$T$, except for the previous $[2/2]$ Canterbury approximant, but, as in the
case of slow-roll Hubble expansions and Canterbury approximants, the center
manifold expansions and Pad{\'e} approximants yield better results for the
attractor solution for large $T$, as can be seen by comparing
Tables~\ref{tab:cexpq0} and~\ref{tab:padeTtheta} with
Table~\ref{tab:phidotrelerr} for $q=0$.
\begin{table}[ht!]
\begin{center}
 \begin{tabular}{|c|c|c|c|c|c|c|c|c|} \hline
                       &             &          &          &          &          &             &          &    \\ [-2ex]
                       &    Num.     &   0th   &   1st   &   2nd   &   [1/1]  &     3rd    &    4th  &   [2/2]    \\ [1ex] \hline
                       &             &          &          &          &          &             &          &    \\ [-2ex]
   $T$                 & 0.6646      & 0.6340   &  0.7101  &  0.6186  &  0.6420  & \textemdash &  0.5676  & \textemdash \\ [1ex] \hline
                       &             &          &          &          &          &             &          &    \\ [-2ex]
$\frac{|\Delta T|}{T}$ & \textemdash & 4.604\%  &  6.846\% &  6.923\% &  3.400\% & \textemdash & 14.602\% & \textemdash  \\ [1ex] \hline
                       &             &          &          &          &          &             &          &    \\ [-2ex]
$\frac{H}{m}$          & 0.5047      & 0.5774   &  0.4083  &  0.6166  &  0.5576  & \textemdash &  0.7619  & \textemdash   \\ [1ex]\hline
                       &             &          &          &          &          &             &          &    \\ [-2ex]
$\frac{|\Delta H|}{H}$ & \textemdash & 14.400\% & 19.111\% & 22.168\% & 10.487\% & \textemdash &  50.969\%& \textemdash \\ [1ex]\hline
\end{tabular}
\end{center}
\caption{Numerical values and relative errors for the various $\dot{\phi}$
slow-roll expansions and Canterbury approximants at
$q=0$.}\label{tab:phidotrelerr}
\end{table}
Among the two classes of slow-roll approximations, the $[1/1]$ slow-roll
Canterbury $\dot{\phi}$ approximant is the best one for large $T$. However,
as can be seen from Figure~\ref{fig:2phidotcomp}, the $[3/3]_\theta$ Pad{\'e}
approximant is a more accurate approximation than the $[1/1]$ Canterbury
$\dot{\phi}$ slow-roll approximation everywhere, while the $[1/1]_\theta$
Pad{\'e} approximant gives a better result for large $|\theta|$ and $T$.
%

\section{Concluding remarks}\label{sec:concl}

In this paper we have used the minimally coupled scalar field with a
quadratic potential in flat FLRW cosmology to illustrate how a global
dynamical system can yield a global understanding of the solution space as
well as the solution's asymptotic features. We have also used center manifold
expansions and Pad{\'e} approximants to obtain approximations for the
attractor solution at early times. In addition we have showed that late time
asymptotics for all solutions is associated with a limit cycle that
demonstrates that the present models exhibit future asymptotic manifest
self-similarity breaking, which physically corresponds to that physical
dimensionless observables such as the deceleration parameter have no future
asymptotic limits, but instead oscillate. This is an issue that becomes even
more pertinent when one includes additional sources such as perfect fluids
and we will return to this in a forthcoming paper dealing with scalar fields
with monomial potentials and perfect fluids. In this case there is a
`competition' between the different matter sources, reminiscent of the
dynamics of Lotka-Volterra systems, as described in~\cite{peretal14}.

In addition we have shown that the slow-roll approximation and expansions are
actually a collage of two types of approximations that describe the early
behavior of the attractor solution. These approximations were subsequently
compared with the center manifold approximations. It was found that center
manifold expansions and associated approximants have a larger range of
convergence than the slow-roll approximation and associated approximants.
Furthermore, it was shown that the $[1/1]_\theta$ Pad{\'e} approximant gives
a better approximation than the slow-roll approximation to the attractor
solution everywhere. The large range of convergence for center manifold
approximants made it possible to join them with approximations describing the
dynamics at late times to produce global approximations for the attractor
solution.

Different convergence properties of various approximation schemes might have
far reaching consequences, so let us therefore take a closer look at the
slow-roll approximation. Expressed in $\theta$ and $T$ the slow-roll
approximation takes the following form for the quadratic scalar field
potential:
\begin{equation}
\theta_\mathrm{SR} = -\arcsin{\left(\frac{T}{3(1-T)}\right)}.
\end{equation}
Performing a Taylor expansion around $T=0$ yields
\begin{equation}
\begin{split}
\theta_\mathrm{SR} &= -\frac13\left\{T + T^2 + \underbrace{\frac{55}{2\cdot3^3}}_{1.019}T^3 + \underbrace{\frac{19}{2\cdot3^2}}_{1.056}T^4 -
\underbrace{\frac{1201}{2^3\cdot3^3\cdot 5}}_{1.112}T^5 + \underbrace{\frac{257}{2^3\cdot3^3}}_{1.190}T^6\right. \\ & \\
& \qquad \left. + \underbrace{\frac{105467}{2^4\cdot3^6\cdot 7}}_{1.292}T^7 +
\underbrace{\frac{16583}{2^4\cdot3^6}}_{1.422}T^8 + \underbrace{\frac{11980259}{2^7\cdot3^{10}}}_{1.585}T^9 +
\underbrace{\frac{7510063}{2^7\cdot3^8\cdot 5}}_{1.789}T^{10}\right\} + {\cal O}(T^{11}).
\end{split}\label{SRTexp}
\end{equation}
A comparison with the center manifold expansion given in
eq.~\eqref{CM_Expansion} shows that it is only the first two terms that
coincide. For every other order the errors for the slow-roll expansion
in~\eqref{SRTexp} are larger than those in~\eqref{CM_Expansion}, and a
similar statement holds for the associated Pad{\'e} approximants. It is only
the exact translations of the slow-roll Hubble expansions and Canterbury
approximants that give competitive results for small $T$ when compared with
the lower order center manifold expansions, which in turn corresponds to
nonlinear variable transformations between $(\phi,H)$ and $(\theta,T)$.

Thus the present paper provides specific examples of how nonlinear
transformations and different approximation schemes significantly affects
convergence rates and ranges for flat FLRW cosmology with a minimally coupled
scalar field with a quadratic potential, which should not come as a surprise.
Nevertheless, this leads to the following questions: Is it possible to find
nonlinear transformations, and other approximation schemes, that yield even
more accurate approximations with a larger range of convergence than the
slow-roll and center manifold expansions and Pad{\'e} and Canterbury
approximants? To what extent does the conclusions we have obtained for the
minimally coupled scalar field with a quadratic potential hold for other
scalar field potentials and other gravity theories? As an aside, it is
interesting to note that somewhat similar issues occur in cosmography, as
exemplified by e.g.~\cite{gruluo14,catvis08}, and references therein. The
above issues are intriguing and we will return to them in future work. In
particular we are going to show that many of the main results and conclusions
in the present paper can be generalized to sources that consist of more
general scalar field potentials and perfect fluids in a forthcoming paper
(both as regards global aspects and how to improve on slow-roll
approximations), although additional, sometimes somewhat surprising, aspects
and phenomena to those discussed in the present paper occur as well.

\subsection*{Acknowledgments}
AA is supported by the projects CERN/FP/123609/2011, EXCL/MAT-GEO/0222/2012,
and CAMGSD, Instituto Superior T{\'e}cnico through FCT plurianual funding,
and the FCT grant SFRH/BPD/85194/2012. Furthermore, AA also thanks the
Department of Engineering and Physics at Karlstad University, Sweden, for
kind hospitality.


\label{bibbegin}


\begin{thebibliography}{99}

\bibitem{bicep2} BICEP2 Collaboration, P.~Ade {\it et al}.
\newblock Detection of B-Mode Polarization at Degree Angular Scales by BICEP2.
\newblock Phys. Rev. Lett. {\bf 112} 241101 (2014).


\bibitem{okaetal14} N.~Okada, V.~N.~Senoguz and Q.~Shafi.
\newblock Simple Inflationary Models in Light of BICEP2: an update.
\newblock arXiv:1403.6403 (2014).


\bibitem{mawan14} Y.~Z.~Ma and Y.~Wang.
\newblock Reconstructing the local potential of inflation with BICEP2 data.
\newblock JCAP {\bf 09} 041 (2014)
\newblock DOI: 10.1088/1475-7516/2014/09/041



\bibitem{beletal85a} V.~A.~Belinski\v{\i}, L.~P.~Grishchuk, Ya. B. Zel'dovich, and I. M. Khalatnikov.
\newblock Inflationary stages in cosmological models with a scalar field.
\newblock  Zh. Eksp. Teor. Fiz. {\bf 89}, 346-360 (1985)


\bibitem{beletal85b} V.~A.~Belinski\v{\i}, L.~P.~Grishchuk, I.~M.~Khalatnikov
    and
    Y.~B.~Zeldovich.
\newblock Inflationary stages in cosmological models with a scalar field.
\newblock Phys.\ Lett.\ {\bf B155} 232 (1985).
\newblock DOI: 10.1016/0370-2693(85)90644-6

\bibitem{belkha87} V.~A.~Belinski\v{\i} and I.~M.~Khalatnikov.
\newblock The degree of generality of inflationary solutions in cosmological models with a scalar field.
\newblock Zh. Eksp. Teor. Fiz. {\bf 93}, 784-799 (1987)

\bibitem{beletal88} V.~A.~Belinski\v{\i}, H.~Ishihara, I.~M.~Khalatnikov and
    H.~Sato.
\newblock On the Degree of Generality of Inflation in Friedmann
Cosmological Models with a Massive Scalar Field.
\newblock Prog.\ Theor.\ Phys.\ {\bf 79} 676 (1988).
\newblock DOI: 10.1143/PTP.79.676

\bibitem{ren02} A.~D.~Rendall.
\newblock Cosmological models and centre manifold theory.
\newblock Gen. Rel. Grav.\ {\bf 34} 1277 (2002).
\newblock DOI: 10.1023/A:1019734703162

\bibitem{ren07} A.~D.~Rendall.
\newblock Late-time oscillatory behaviour for self-gravitating scalar fields.
\newblock Class.\ Quant.\ Grav.\ {\bf 24} 667 (2007).
\newblock DOI: 10.1088/0264-9381/24/3/010	

\bibitem{lidetal94} A.~R.~Liddle, P.~Parsons and J.~D.~Barrow.
\newblock Formalizing the slow-roll approximation in inflation.
\newblock Phys.\ Rev.\ {\bf D50} 7222 (1994).
\newblock DOI: 10.1103/PhysRevD.50.7222

\bibitem{col03} A.~A.~Coley.
\newblock {\em Dynamical systems and cosmology}.
\newblock Kluwer Academic Publishers, Dordrecht, (2003).

\bibitem{beyesc13} F.~Beyer and L.~Escobar.
\newblock Graceful exit from inflation for minimally coupled Bianchi A scalar field
models.
\newblock Class.\ Quant.\ Grav.\ {\bf 30} 195020 (2013).
\newblock DOI: 10.1088/0264-9381/30/19/195020

\bibitem{fadetal14} C.~R.~Fadragas, G.~Leon and E.~N.~Saridakis.
\newblock Dynamical analysis of anisotropic scalar-field cosmologies for a wide range
of potentials.
\newblock  Class.\ Quant.\ Grav.\ {\bf 31} 075018 (2014).
\newblock DOI: 10.1088/0264-9381/31/7/075018.

\bibitem{ugg13b} C.~Uggla.	
\newblock Recent developments concerning generic spacelike singularities.
\newblock Gen.\ Rel.\ Grav.\ {\bf 45} 1669 (2013). 
\newblock DOI: 10.1007/s10714-013-1556-3

\bibitem{waiell97} J.~Wainwright and G.~F.~R.~Ellis.
\newblock {\em Dynamical systems in cosmology}.
\newblock {C}ambridge {U}niversity {P}ress, Cambridge, (1997).

\bibitem{cra91} J.~D.~Crawford.
\newblock Introduction to bifurcation theory.
\newblock Rev.\ Mod.\ Phys. {\bf 63} 991 (1991). 

\bibitem{car81} J.~Carr.
\newblock {\em Applications of center manifold theory}.
\newblock Springer Verlag, New York, 1981.

\bibitem{remcar13} G.~N.~Remmen and S.~M.~Carroll.
\newblock Attractor solutions in scalar-field cosmology.
\newblock Phys.\ Rev.\ {\bf D88} 083518 (2013).
\newblock DOI: 10.1103/PhysRevD.88.083518

\bibitem{remcar14} G.~N.~Remmen and S.~M.~Carroll.
\newblock How many $e$-folds should we expect from high-scale inflation?
\newblock Phys.\ Rev.\ {\bf D90} 063517 (2014).
\newblock DOI: 10.1103/PhysRevD.90.063517

\bibitem{corslo14} A.~Corichi and D.~Sloan.
\newblock Inflationary Attractors and their Measures.
\newblock Class.\ Quant.\ Grav.\ {\bf 31} 062001 (2014).
\newblock DOI: 10.1088/0264-9381/31/6/062001

\bibitem{ugg13} C.~Uggla.
\newblock Global cosmological dynamics for the scalar field representation of the modified Chaplygin
gas.
\newblock Phys. Rev. {\bf D88} (2013) 064040.
\newblock DOI: 10.1103/PhysRevD.88.064040

\bibitem{tur83} M.~S.~Turner.
\newblock Coherent scalar-field oscillations in an expanding universe.
\newblock Phys. Rev.\ {\bf D28} 1243 (1983).
\newblock DOI: 10.1103/PhysRevD.28.1243

\bibitem{dammuk98} T.~Damour and V.~F.~Mukhanov.
\newblock Inflation without Slow Roll.
\newblock Phys.\ Rev.\ Lett.\ {\bf 80} 3440 (1998).
\newblock DOI: 10.1103/PhysRevLett.80.3440

\bibitem{macpic00} A.~de la Macorra and G.~Piccinelli.
\newblock General scalar fields as quintessence.
\newblock Phys.\ Rev.\ {\bf D61} 123503 (2000).
\newblock DOI: 10.1103/PhysRevD.61.123503

\bibitem{waietal99} J.~Wainwright, M.~J.~Hancock and C.~Uggla.
\newblock Asymptotic self-similarity breaking at late times in cosmology.
\newblock Class.\ Quant.\ Grav.\ {\bf 16} 2577 (1999).
\newblock DOI: 10.1088/0264-9381/16/8/302

\bibitem{fos98} S.~Foster.
\newblock Scalar Field Cosmologies and the Initial Space-Time Singularity.
\newblock Class.\ Quant.\ Grav.\ {\bf 15} 3485 (1998).

\bibitem{sinetal03}  J.~J.~Sinou, F.~Thouverez and L.~Jezequel  .
\newblock Extension of the center manifold approach, using rational fractional
approximants, applied to non-linear stability analysis.
\newblock Nonlinear Dynamics\ {\bf 33} 267  (2003).
\newblock DOI: 10.1023/A:1026060404109

\bibitem{bohetal12} C.~ G.~B{\"oh}mer, N.~Chan, and R.~Lazkoz.
\newblock Dynamics of dark energy models and centre manifolds.
\newblock Physics Letters B, {\bf 714}, 11 (2012). 
\newblock DOI: 10.1016/j.physletb.2012.06.064.

\bibitem{bak75} G.~A.~Baker.
\newblock {\em Essentials of Pad\'e Approximants}.
\newblock {A}cademic {P}ress, New York, (1975).

\bibitem{kal02} J.~Kallrath.
\newblock {\em On Rational Function Techniques and Pad{\'e} Approximants. An
Overview.}
\newblock (2002).

\bibitem{gruluo14} C.~Gruber and O.~Luongo.
\newblock Cosmographic analysis of the equation of state of the universe through Pad\'e approximations.
\newblock Phys.\ Rev.\ {\bf D89} 103506 (2014).
\newblock DOI: 10.1103/PhysRevD.89.103506

\bibitem{weietal14} H.~Wei, X.~P.~Yan and Y.~N.~Zhou.
\newblock Cosmological applications of Pad\'e approximant.
\newblock JCAP\ {\bf 2014} 045 (2014).
\newblock DOI: 10.1088/1475-7516/2014/01/045

\bibitem{ugg89} C.~Uggla.
\newblock Asymptotic cosmological solutions: orthogonal Bianchi type II models.
\newblock Class.\ Quant.\ Grav.\ {\bf 6} 383 (1989).

\bibitem{limetal04} W.~ C.~Lim, H.~van Elst, C.~ Uggla, and J.~Wainwright.
\newblock Asymptotic isotropization in inhomogeneous cosmology.
\newblock Phys.\ Rev.\ {\bf D69}~:~103507 (2004).

\bibitem{ven14} V.~Vennin.
\newblock Horizon-Flow off-track for Inflation.
\newblock Phys.\ Rev.\ {\bf D89} 083526 (2014).
\newblock DOI: 10.1103/PhysRevD.89.083526.

\bibitem{salbon90} D.~S.~Salopek and J.~R.~Bond.
\newblock Nonlinear evolution of long-wavelength metric fluctuations in
inflationary models.
\newblock Phys.\ Rev.\ {\bf D42} 3936 (1990).

\bibitem{peretal14} J.~Perez, A.~F{\"u}zfa, T.~Carletti, L.~M{\'e}lot,
    L.~Guedezounme.
\newblock The Jungle Universe.	
\newblock Gen. Rel. Grav.\ {\bf 46} 1753  (2014).
\newblock DOI: 10.1007/s10714-014-1753-8

\bibitem{catvis08} C.~Cattoen and M.~Visser.
\newblock Cosmographic Hubble fits to the supernova data.
\newblock Phys.\ Rev.\ {\bf D78} 063501 (2008).
\newblock DOI: 10.1103/PhysRevD.78.063501




















\end{thebibliography}
\end{document}